\documentclass[12pt]{article}
\usepackage{latexsym}
\usepackage{amsmath,,calrsfs}
\usepackage{amsfonts}
\usepackage{amssymb}
\usepackage{amscd}
\usepackage{fancybox}
\usepackage{cite}
\usepackage{amsmath,amsfonts,amsbsy}
\usepackage{pstricks,pst-node}
\usepackage[small,bf,hang]{caption2}
\usepackage{graphicx}
\usepackage{epsfig}
\usepackage{psfrag}

\usepackage{float}

\psset{unit=1.3cm,linewidth=.5pt,radius=.2}  

\usepackage{multirow}                     
\usepackage{float}                          
\usepackage{lscape}                         
\usepackage{bm}

\newcommand{\bb}{\bar\beta}

\newcommand{\beq}{\begin{equation}}
\newcommand{\eeq}{\end{equation}}
\newcommand{\bi}{\begin{itemize}}
\newcommand{\ei}{\end{itemize}}
\newcommand{\bt}{\begin{tabular}}
\newcommand{\et}{\end{tabular}}
\newcommand{\bc}{\begin{center}}
\newcommand{\ec}{\end{center}}

\newcommand{\ft}[2]{{\textstyle {\frac{#1}{#2}} }}

\newcommand{\be}{\begin{equation}}
\newcommand{\ee}{\end{equation}}
\newcommand{\bea}{\begin{eqnarray}}
\newcommand{\eea}{\end{eqnarray}}
\newcommand{\ba}{\begin{array}}
\newcommand{\ea}{\end{array}}

\def\bbox{{\,\lower0.9pt\vbox{\hrule \hbox{\vrule height 0.2 cm
\hskip 0.2 cm \vrule height 0.2 cm}\hrule}\,}}
\newcommand{\dsl}{\pa \kern-0.5em /}

\newcommand{\nn}{\nonumber \\}

\font\mybb=msbm10 at 12pt
\def\bb#1{\hbox{\mybb#1}}

\def\bR {\bb{R}}

\makeatletter \@addtoreset{equation}{section} \makeatother

\def\slashchar#1{\setbox0=\hbox{$#1$}           
   \dimen0=\wd0                                 
   \setbox1=\hbox{/} \dimen1=\wd1               
   \ifdim\dimen0>\dimen1                        
      \rlap{\hbox to \dimen0{\hfil/\hfil}}      
      #1                                        
   \else                                        
      \rlap{\hbox to \dimen1{\hfil$#1$\hfil}}   
      /                                         
   \fi}

\def\eq#1{(\ref{#1})}

\begin{document}

\begin{titlepage}
\begin{center}

\hfill UG-10-18 \\ \hfill MIT-CTP-4129\\ \hfill DAMTP-2010-15\\ \hfill MIFP-10-8

\vskip 1.5cm

{\Large \bf  More on Massive 3D Supergravity}

\vskip 1cm

{\bf Eric A.~Bergshoeff\,$^1$, Olaf Hohm\,$^2$, Jan Rosseel\,$^1$ \\[0.5ex] Ergin Sezgin\,$^3$ and Paul K.~Townsend\,$^4$} \\

\vskip 25pt

{\em $^1$ \hskip -.1truecm Centre for Theoretical Physics, University of Groningen, \\ Nijenborgh 4, 9747 AG Groningen, The Netherlands \vskip 5pt }

{email: {\tt E.A.Bergshoeff@rug.nl, j.rosseel@rug.nl}} \\

\vskip 15pt

{\em $^2$ \hskip -.1truecm Center for Theoretical Physics, Massachusetts Institute of Technology, \\ Cambridge, MA 02139, USA \vskip 5pt }

{email: {\tt ohohm@mit.edu}} \\

\vskip 15pt

{\em $^3$ \hskip -.1truecm George and Cynthia Woods Mitchell Institute for Fundamental Physics and Astronomy, Texas A\& M University, College Station,
TX 77843, USA}

{email: {\tt sezgin@tamu.edu}}

\vskip 15pt

{\em $^4$ \hskip -.1truecm Department of Applied Mathematics and Theoretical Physics,\\ Centre for Mathematical Sciences, University of Cambridge,\\
Wilberforce Road, Cambridge, CB3 0WA, U.K. \vskip 5pt }

{email: {\tt P.K.Townsend@damtp.cam.ac.uk}} \\

\end{center}

\vskip 0.5cm

\begin{center} {\bf ABSTRACT}\\[3ex]
\end{center}

Completing earlier work on three dimensional (3D) ${\cal N}=1$ supergravity with curvature-squared terms, we construct the  general supergravity extension of `cosmological'  massive gravity theories.  In particular, we show that all  adS vacua of ``new massive gravity'' (NMG) correspond to supersymmetric adS vacua of
a  ``super-NMG''  theory that  is perturbatively unitary whenever the corresponding NMG theory is perturbatively unitary.








\end{titlepage}

\newpage
\setcounter{page}{1}
\tableofcontents

\newpage


\section{Introduction}\setcounter{equation}{0}

The local dynamics of Einstein's general relativity for a three-dimensional spacetime is trivial because Einstein's equations imply that the spacetime
curvature is zero in the absence of sources  \cite{Staruszkiewicz:1963zz,Leutwyler,Deser:1983tn}. The addition to the standard Einstein-Hilbert (EH) action of
curvature-squared terms leads to non-trivial  dynamics but, typically, some propagated modes have negative energy, implying ghost particles in the
quantum theory and a corresponding loss of unitarity.  This is an inevitable feature in four spacetime dimensions \cite{Stelle:1976gc} but it was
recently  discovered \cite{Bergshoeff:2009hq} that ghosts can be avoided in  three dimensions (3D) if (i) the EH term has the `wrong' sign  and (ii)
the curvature-squared invariant is constructed from the scalar\footnote{See also the discussion in \cite{Bergshoeff:2009tb,Bergshoeff:2009fj}.}
\be\label{Kinv} K=
R^{\mu\nu}R_{\mu\nu} - \frac{3}{8} R^2\, , \ee where $R_{\mu\nu}$ is the Ricci tensor,  and $R$ its trace, for a metric $g$ which we take to have
`mostly plus' signature. An equivalent expression is $K= G^{\mu\nu}S_{\mu\nu}$, where $G_{\mu\nu}$ is the Einstein tensor and $S_{\mu\nu}$ the
Schouten tensor (the 2nd order `potential' for the 3rd order Cotton tensor, which is the 3D analog of the Weyl tensor). The inclusion of this K-term
in the action introduces a mass parameter $m$ and linearizing about the Minkowski vacuum one finds that  two modes of helicities\footnote{We use  ``helicity''  to mean  ``relativistic helicity'', i.e.  the scalar product of  the {\it relativistic} 3-momentum with the Lorentz rotation 3-vector, divided by the mass.}
$\pm2$ are
propagated, unitarily,  with mass $m$. This model is now generally referred to as  ``new massive gravity'' (NMG).   The addition of a (parity
violating) Lorentz Chern-Simons (LCS) term leads to a model   that  propagates the helicity $\pm2$ modes with different masses $m_\pm$
\cite{Bergshoeff:2009hq}; this has been called ``general massive gravity'' (GMG).   The limit of GMG in which $m_-\to\infty$ for fixed $m_+$ yields
the well-known   ``topological massive gravity'' (TMG) \cite{Deser:1981wh}.

All these models have `cosmological' extensions in which a cosmological constant term is added to the Lagrangian density; we may take this to be
$-2m^2\lambda$ times the volume density, where $\lambda$ is a {\it dimensionless} cosmological parameter. In this context it is convenient to allow for an
arbitrary coefficient $\sigma$ of  the EH term, so the Lagrangian density for cosmological GMG is \be {\cal L}_{GMG} = \sqrt{-\det g} \left [
-2\lambda m^2 + \sigma R + \frac{1}{m^2} K \right] + \frac{1}{\mu} {\cal L}_{LCS}\, , \ee where ${\cal L}_{LCS}$ is the Lorentz-Chern-Simons density.
When   $\lambda=0$ there is a Minkowski vacuum in which are propagated two modes,  of helicities $+2$ and $-2$, and these are propagated unitarily  as
long as $\sigma<0$ and $m^2>0$; for
$\sigma=-1$ this is the GMG model described above, with masses $m_\pm$ such that $m^2= m_+m_-$ and $\mu = m_+ m_-/(m_--m_+)$.
More generally, it is convenient to allow for either sign of $m^2$, in addition to either sign of $\sigma$, because one does not know, a priori, what
unitarity will permit in non-Minkowski vacua.  Note, however, that a  change in sign of both $\sigma$ and $m^2$  is equivalent to a change in the
overall  sign of the $\mu$-independent terms in the action, from which it  follows that the dependence of the field equations  on the signs of  $\sigma$ and $m^2$ is entirely
through the sign of  the product $m^2\sigma$. The same is true of the space of solutions, in particular vacuum solutions, although conclusions
concerning the unitarity of modes propagated in a given vacuum will depend on the individual signs of both $\sigma$ and $m^2$.

All maximally-symmetric vacua of GMG were found in
\cite{Bergshoeff:2009hq}. By definition, such vacua have the
property that \be\label{maxsym} G_{\mu\nu} = -\Lambda g_{\mu\nu}\;,
\ee where $\Lambda$ is the cosmological constant, which is positive
for de Sitter (dS) vacua and negative for anti-de Sitter (adS)
vacua, and zero for Minkowski vacua. When curvature-squared terms
are present it is important to distinguish the cosmological constant
$\Lambda$ from the cosmological parameter $\lambda$, which  becomes
a quadratic  function of $\Lambda$: \be\label{lamLam1} 4m^4 \lambda
= \Lambda \left(\Lambda + 4m^2 \sigma\right)\, . \ee Observe that
zero cosmological {\it term} allows non-zero  cosmological {\it
constant}; this is a typical feature of higher-derivative gravity
theories first pointed out in \cite{Boulware:1985wk}. Of particular
interest  in the present context are the adS vacua because of their
possible association with a holographically dual conformal field
theory (CFT) via the ${\rm adS}_3/{\rm CFT}_2$ correspondence
\cite{Brown:1986nw,Maldacena:1997re}.  In this connection, it was
shown for NMG in \cite{Bergshoeff:2009aq} (completing earlier
partial results \cite{Liu:2009pha})  that  the boundary CFT is
non-unitary whenever  the `bulk'  gravity theory is unitary, and
vice-versa, although there is a special case (recently analyzed in
more detail
\cite{Grumiller:2009sn,Hohm:2010jc,Alishahiha:2010bw,Grumiller:2008qz})
in which the central charge vanishes and the bulk massive gravitons
are replaced by bulk massive `photons'.  This result was
disappointing, but perhaps to be expected in light of the similar
difficulty afflicting cosmological TMG (we refer the reader to
\cite{Grumiller:2009mw,Skenderis:2009nt,Maloney:2009ck,Bergshoeff:2010iy}
for up-to-date accounts). An obvious question is whether this
situation is any different in the context of a supergravity
extension of GMG.

The off-shell ${\cal N}=1$ `graviton' supermultiplet \cite{Howe:1977ut,Uematsu:1984zy} comprises the dreibein (from which one constructs the metric),  the 3D Rarita-Schwinger potential and a scalar field $S$. The off-shell supersymmetry transformations are independent of the choice of action and it is possible to determine the general supersymmetric field configuration without reference to the action \cite{Andringa:2009yc}. In particular, a maximally symmetric vacuum is supersymmetric provided that \be\label{SLam} S^2=-\Lambda\, , \ee which is, of course,  possible only when $\Lambda\le0$, i.e. for Minkowski or adS vacua.  In the absence of the supergravity cosmological term, which is proportional to $S$, one does not need the details of the non-linear theory to see that  $S=0$ is a solution of the field equation for $S$, and hence that there exists a supersymmetric Minkowski vacuum.  The general conditions for unitarity of the linear theory in this vacuum were obtained in \cite{Andringa:2009yc}, extending an analysis applied earlier to NMG  \cite{Deser:2009hb}.  Generically, the scalar field $S$ has a kinetic term,  and  there is one unitary model of this type:  the supersymmetric extension of the $R+R^2$ model.  Otherwise,   unitarity in the Minkowski vacuum requires that $S$ be ``auxiliary'', in the sense that there is no $\left(\partial S\right)^2$ term, and this is indeed the case  for any supersymmetric extension of GMG, as was established  already in \cite{Bergshoeff:2009hq}  by adapting earlier general results \cite{Nishino:2006cb}.

A fully non-linear ${\cal N}=1$ 3D supergravity model with generic curvature-squared terms was constructed
in \cite{Andringa:2009yc}. This was partly motivated  by the fact that  the non-linear details are crucial  to an understanding of the physics in adS vacua.  One question of obvious interest is whether a given adS vacuum of GMG is supersymmetric in the context of a supergravity extension of GMG. However,  this  question was not answered by the construction of  \cite{Andringa:2009yc}. For the question to make sense one needs  a supergravity model that has (cosmological) GMG as its bosonic truncation after elimination of any auxiliary fields, and it is implicit in the results of \cite{Andringa:2009yc} that, apparently, there is no such  model!  There is no difficulty in the absence of curvature-squared terms; the EH invariant includes an $S^2$ term and eliminating $S$ converts the supergravity cosmological term proportional to $S$ into a standard cosmological term allowing (supersymmetric) adS vacua. However, the supersymmetric extension of the NMG curvature-squared scalar $K$
presented in \cite{Andringa:2009yc} includes both an $S^4$ and an $RS^2$ term, so the $S$ equation of motion is now cubic with $R$-dependent coefficients.  Elimination of $S$ then leads to an  infinite power series in $R$ (irrespective of the ambiguity in the choice of  solution to a cubic equation). This means  that none of the supergravity models constructed in  \cite{Andringa:2009yc} can really be considered to be a ``super-GMG'' model,  except in the super-TMG limit  (which has been known for some
time  \cite{Deser:1982sw,Deser:1982sv,vanNieuwenhuizen:1985cx}).

This state of affairs suggests that there was some ingredient missing from the analysis of  \cite{Andringa:2009yc}.  In this paper we supply the missing ingredient, and this  allows an analysis of  unitarity for  massive supergravity theories in adS vacua.  The crucial observation is that there is an additional super-invariant that includes both $RS^2$ and $S^4$ terms but {\it no curvature-squared term}. This was missed in
\cite{Andringa:2009yc} because that paper only aimed to construct a supersymmetric extension of the $K$ and $R^2$ invariants; this was achieved but without the appreciation that the result is not unique. Taking into account the new super-invariant, one can find a supersymmetrization of the $K$ invariant that includes an $S^4$ term but {\it not} an $RS^2$ term\footnote{Or vice versa. As already observed in \cite{Bergshoeff:2009hq},  one of the two {\it must} be present because $S$ can be entirely absent only from  super-conformal invariants.}. There is a similar new invariant that can contribute  at the same dimension as the LCS term; although it includes an apparently undesirable $RS$ term, its effects may cancel against those of the $RS^2$ term for special values of $S$. This possibility motivates us to start with the most general model containing  no terms of dimension higher than $R^2$ but all terms of this dimension or less. This general supergravity model contains two additional mass parameters as compared with the model constructed in \cite{Andringa:2009yc}.

Of most interest are those special cases of the general model for which $S$ can be eliminated by an algebraic equation with constant coefficients; in such cases, the bosonic truncation yields a model of precisely GMG type.  As will become clear, there is a  simple
subclass of such models, which we refer to collectively as ``super-GMG'', that is  parametrized by the same two mass parameters $(m,\mu)$ as GMG itself.   It turns out that not all maximally symmetric vacua of GMG are solutions of super-GMG; some dS vacua are excluded. In contrast, all adS vacua of GMG continue to be solutions of super-GMG, although some map to {\it two} adS vacua of super-GMG because the latter are distinguished by their dependence on a  cosmological mass parameter  $M$ that differs from (and is non-linearly related to) the cosmological parameter $\lambda$ of GMG.  This result  allows us to address the question of which adS vacua of GMG are supersymmetric solutions of super-GMG. What we find can be summarized  by saying that {\it all adS vacua of GMG are supersymmetric vacua of super-GMG but super-GMG has additional adS vacua that are not supersymmetric}.

Given a vacuum solution, the next step is to determine the quadratic approximation to the action linearized about it, and thence the
nature of the modes propagated, in particular whether they  are physical or ghosts.  This settles the issue of {\it perturbative} unitarity.
Perturbative unitarity is a necessary condition for unitarity, and may be sufficient in Minkowski vacua, but it is  not  sufficient in adS vacua
because there are then non-perturbative excitations to take into account; {\it viz}. BTZ black holes. In the context of TMG there is the, by now
well-known,  problem that the `wrong-sign'  of the EH term needed for perturbative unitarity implies a negative mass for BTZ black holes, which
translates  to a negative central charge of the boundary CFT, although it has been suggested that  a superselection principle may allow the
consistent exclusion of  BTZ black holes \cite{Deser:2010df}. In any case, we limit ourselves in this paper to a discussion of perturbative unitarity.

In the supergravity context an analysis of perturbative unitarity
generally requires an analysis of fermionic field fluctuations, as
well as bosonic field fluctuations, but  supersymmetric vacua are
exceptional because  perturbative unitarity of the bosonic
fluctuations implies perturbative unitarity of the fermionic
fluctuations. This feature  of supersymmetric vacua greatly
simplifies the analysis, and for this reason we consider here only
supersymmetric vacua. The results of \cite{Andringa:2009yc} for the
supersymmetric Minkowski vacuum are still valid for the larger class
of supergravity models found here, for reasons already explained, so
that leaves the supersymmetric adS vacua. A complete analysis of
perturbative unitarity for the adS vacua of NMG was presented in
\cite{Bergshoeff:2009aq}. No analogous analysis for supergravity was
attempted in  \cite{Andringa:2009yc}, mainly because  of the
problems already mentioned with the model constructed there. Here we
shall show how the analysis of \cite{Bergshoeff:2009aq} for
perturbative unitarity of NMG extends to the supersymmetric adS
vacua of super-NMG. In particular, we shall show that the super-NMG
model is perturbatively unitary in a supersymmetric adS vacuum
whenever the corresponding NMG model is perturbatively unitary.

This paper is organized as follows. In section \ref{sec:invariants} we determine the new
super-invariants by means  of the superconformal approach. These are then used in
section \ref{sec:genmod} to construct the bosonic truncation of the general curvature-squared supergravity model,
in which context we determine all  maximally-symmetric vacua and revisit  pp-wave solutions.
In section \ref{sec:superGMG} we specialize to models in which the scalar field $S$ is ``auxiliary'' in the sense
explained above. It turns out that  this condition still allows propagating fluctuations of  $S$; we refer to those cases in which this does {\it not}
happen as  ``generalized super-GMG'' and it is in this context that we find the``super-GMG'' models  that have GMG as a bosonic truncation. In section \ref{sec:bulkunitarity} we further specialize to super-NMG,  and its ``generalized'' extension, determining the conditions for perturbative
unitarity in supersymmetric adS vacua.  We present our conclusions, with some further discussion,   in section \ref{sec:conclusions}.


\section{3D supergravity invariants}\setcounter{equation}{0}
\label{sec:invariants}


In order to determine the  bosonic terms of  3D supergravity actions involving curvature squared terms, it is convenient to combine global
supersymmetry with local conformal symmetry. In the  conformal approach one first constructs a superconformal gauge invariant action involving  one or
more compensating multiplets, which are then used to gauge fix the superfluous superconformal symmetries to arrive at  a  standard Poincar\'e
supergravity invariant.  For our purposes, we do not need to perform the complete conformal programme. We only need to construct globally
supersymmetric actions that can be made invariant under local conformal transformations. This is because global supersymmetry
connects the $S$-dependent terms in the action to the (possibly higher-derivative)  kinetic terms for the compensating supermultiplet, and
local conformal invariance connects these kinetic terms to the $R$-dependent terms. After fixing the compensating fields one ends up with an action containing all relevant $R^2$ and $S$-dependent terms. The results are consistent with the bosonic truncations  of the super-invariants found in \cite{Andringa:2009yc} but, surprisingly, we also find the bosonic truncation of  a new super-invariant. We will begin by recalling the essentials  of the conformal procedure and then show how the bosonic truncations of all relevant super-invariants may be determined.


\subsection{${\cal N} = 1$ superconformal tensor calculus}


One starts with a (globally) supersymmetric action, involving one or more compensating multiplets. These can then be coupled to the conformal
supergravity multiplet, that consists of the dreibein $e_\mu{}^a$ and the gravitino $\psi_\mu$, with the following transformation rules under
fermionic symmetries: \be \label{confsugra} \delta e_\mu{}^a =  \frac{1}{2}\bar\epsilon\gamma^a\psi_\mu \, , \qquad \delta\psi_\mu =
D_\mu(\omega)\epsilon +\gamma_\mu\eta\,, \ee where $\epsilon$ is the ordinary $Q$-supersymmetry parameter and $\eta$ is the parameter of the special
$S$-supersymmetries.

In the following we will be mainly interested in the bosonic part of the action. Restricting our attention to the bosonic level, conformal invariance
means invariance under dilatations $D$ and special conformal transformations $K_a$. Invariance of a Lagrangian under these transformations can be
achieved in three steps:
\begin{itemize}
\item In a first step, one ensures that all terms in the Lagrangian have the correct behavior under global dilatations. Under these scale
    transformations, a field $\phi$ transforms with a certain weight $w_\phi$:
\begin{equation}
\delta_D \phi = w_\phi \zeta \phi \,,
\end{equation}
where $\zeta$ denotes the parameter of the dilatations. Invariance of the action under global scale transformations is then accomplished when
the sum of the weights of all fields in each term adds up to the space-time dimension $d$ (where derivatives $\partial_\mu$ have weight one).
\item In a second step, one takes care of the invariance of the action under local dilatations by introducing a gauge field $b_\mu$ that
transforms as follows:
\begin{equation}
\delta_D b_\mu = \partial_\mu \zeta \,.
\end{equation}
All derivatives can then be turned into dilatation-covariant derivatives. E.g. for a field $\phi$ with weight $w_\phi$ this implies the following
substitution:
\begin{equation}
\partial_\mu \phi \ \to \ D_\mu \phi = (\partial_\mu - w_\phi b_\mu)\phi \,.
\end{equation}
In a similar manner one can  replace $\Box \phi$ by a dilatation-covariant expression $\Box^C \phi$:
\begin{equation}
\Box^C \phi = \eta^{ab} D_a D_b \phi = e^{a\mu} \left(\partial_\mu D_a \phi - (w_\phi + 1) b_\mu D_a \phi + \omega_{\mu \, ab} D^b \phi \right)
\,.
\end{equation}
\item In the last step, one takes care of the invariance under special conformal transformations $K_a$. This can be achieved by adding terms
    involving the Ricci tensor and scalar and by taking into account the following transformation rules under $K_a$:
\begin{eqnarray} \label{Ktransf}
\delta_K b_\mu & = & 2 \Lambda_{K\mu}\,, \nonumber \\ \delta_K D_a \phi & = & -2 w_\phi \Lambda_{Ka} \phi \,, \nonumber \\ \delta_K \Box^C \phi &
= & -2 w_\phi (D^c \Lambda_{Kc}) \phi + 2 (d-2-2 w_\phi) \Lambda^c_K D_c \phi \,, \nonumber \\ \delta_K R_{ab} & = & -2 \eta_{ab} D_c \Lambda^c_K
- 2 (d-2) D_a \Lambda_{Kb} \,, \nonumber \\ \delta_K R & = & -4 (d-1) D^c \Lambda_{Kc} \,,
\end{eqnarray}
where $\Lambda_{Ka}$ are the parameters of the special conformal transformations. The fact that $b_\mu$ transforms with a shift under the special
conformal transformations means that, writing out all covariant derivatives, one finds that the dilatation gauge field drops out in any conformal
action.
\end{itemize}
These three steps are  enough to ensure invariance under conformal transformations. In particular, the last step allows one to extract the dependence
of the conformal Lagrangian on the curvatures. By employing a suitable gauge fixing, the (bosonic) Lagrangian invariant under local super-Poincar\'{e}
transformations can then be extracted.

In order to discuss this gauge fixing in more detail, let us note that in the following we will always use an off-shell $\mathcal{N}=1$ scalar
multiplet as compensating multiplet. This consists of a real scalar $\phi$, a Majorana fermion $\lambda$ and a real auxiliary scalar $S$. The
transformation rules under ordinary and special supersymmetry are then given by
\begin{eqnarray} \label{confscal}
\delta\phi &=& \frac{1}{4}\bar\epsilon\lambda\, , \qquad \delta S =  -\bar\epsilon D\hskip -.25truecm \slash \,\lambda
-2(w_\phi-1)\,\bar\lambda\eta\,, \nonumber \\ \delta\lambda &=&  D\hskip -.25truecm \slash \,\phi\,\epsilon-\frac{1}{4}S\epsilon -2w_\phi\phi\,\eta\,
.
\end{eqnarray}
We choose the following gauge fixing conditions:
\begin{eqnarray} \label{gaugefixcond}
K_a-\mathrm{gauge} \quad & : & \quad b_\mu = 0 \,, \nonumber \\ D-\mathrm{gauge} \quad & : & \quad \phi = \phi_0 = \mathrm{constant} \,,  \nonumber\\
S-\mathrm{gauge} \quad & : & \quad \lambda = 0  \, .
\end{eqnarray}
As the $S$-gauge is not invariant under supersymmetry, the
super-Poincar\'{e} rules will involve a compensating
$S$-transformation, with parameter
\begin{equation}
\eta = -\frac{1}{8} \frac{S}{w_\phi \phi_0} \epsilon \,.
\end{equation}
In the following, we will always choose $\phi_0$ such that\footnote{This convention is such that according to (\ref{confsugra}) the final
supersymmetry rule of the gravitino is given by : $\delta \psi_\mu = D_{\mu} (\omega) \epsilon + \frac{1}{2} S \gamma_\mu \epsilon$, as used in
\cite{Andringa:2009yc}.}
\begin{equation} \label{wphi}
w_\phi \phi_0 = -\frac{1}{4} \,.
\end{equation}

Let us illustrate this procedure by constructing the ordinary two-derivative $\mathcal{N}=1$, 3D super-Poincar\'{e} action. We start from the
(globally supersymmetric) action
\begin{equation} \label{Lpoincrigid}
L_{\rm EH}^{\rm rigid} = \phi \Box \phi - \frac{1}{4} \bar{\lambda} \gamma^\mu \partial_\mu \lambda + \frac{1}{16} S^2 \,.
\end{equation}
\emph{}From now on, we will concentrate on the bosonic terms only.
The action corresponding to the Lagrangian (\ref{Lpoincrigid}) is
not yet invariant under local conformal transformations. In order to
render it conformally invariant, we first note that it is invariant
under global scale transformations. These transformations consist of
a scaling of the coordinates and a scaling of the fields according
to the following weights:
\begin{equation}
w_\phi = \frac{1}{2} \,, \quad w_\lambda = w_\phi + \frac{1}{2} = 1 \,, \quad w_S = w_\phi + 1 = \frac{3}{2} \,.
\end{equation}
One then has to replace the derivatives by covariant ones and add extra terms involving curvatures. Using the rules (\ref{Ktransf}), one can check
that the action corresponding to
\begin{equation}\label{Lpoincconf}
L_{\rm EH}^{\rm conf} = -32 \phi \Box^C \phi -2 S^2 + 4 R \phi^2
\end{equation}
is conformally invariant, provided the metric transforms as usual with weight $-2$.
The super-Poincar\'{e} theory can now easily be recovered by using the gauge fixing conditions (\ref{gaugefixcond}) with, as
a consequence of (\ref{wphi}),
\begin{equation}
\phi_0 = -\frac{1}{2} \,.
\end{equation}
One thus finds the following Lagrangian
\begin{equation} \label{Lpoinc}
L_{\rm EH} = R - 2 S^2 + (\mathrm{fermionic}\ \mathrm{terms})\,,
\end{equation}
which is a standard result  \cite{Howe:1977ut}. We next consider a curvature squared term.


\subsection{A supersymmetric curvature squared action}


One can employ a similar reasoning as above starting from the higher-derivative supersymmetric action
\begin{equation} \label{Lrrrigid}
L_{\rm Ric}^{\rm rigid} = \Box \phi \Box \phi - \frac{1}{4} \bar{\lambda} \gamma^\mu \partial_\mu \Box \lambda + \frac{1}{16} S \Box S \,.
\end{equation}
To ensure conformal invariance, one now has to choose different weights:
\begin{equation}
w_\phi = -\frac{1}{2} \,, \quad w_\lambda = w_\phi + \frac{1}{2} = 0 \,, \quad w_S = w_\phi + 1 = \frac{1}{2} \,.
\end{equation}
One can again replace all derivatives by covariant ones and add terms involving the curvatures to obtain a conformally invariant action. Focusing on
the bosonic terms, one obtains the following result:
\begin{equation} \label{Lrrconf}
L_{\rm Ric}^{\rm conf} = 4 \left(\Box^C \phi\right)^2  + \frac{1}{4}S\Box^C S  + 4 \phi^2\left[R^{\mu\nu} R_{\mu\nu} -\frac{23}{64}R^2\right] -\frac{1}{32}S^2
R\,.
\end{equation}
Note that we have only written the relevant bosonic terms in this Lagrangian. The full result contains extra terms\footnote{Of the form $R^{ab} D_a \phi D_b \phi$, $R \phi \Box^C \phi$ and $R (D \phi)^2$.} that vanish upon using the gauge
fixing condition (\ref{gaugefixcond}). The third term
cancels the $K_a$-variation of the $(\Box^C \phi)^2$ term, while the last term cancels the $S \Box^C S$ variation. Upon using the gauge fixing
condition
\begin{equation}
\phi_0 = \frac{1}{2} \,,
\end{equation}
one finds that
\begin{equation} \label{confRsquare}
L_{\rm Ric} = R^{\mu\nu} R_{\mu\nu}-\frac{23}{64}R^2 + \frac{1}{4} S \Box S - \frac{1}{32} S^2 R + (\mathrm{fermionic}\ \mathrm{terms})\,.
\end{equation}


\subsection{A new supersymmetric  $S^n$ action }


An indication for the existence of a new supersymmetric invariant can be obtained by comparing $L_{\rm Ric}$ constructed above with the following two
supersymmetric invariants  constructed in \cite{Andringa:2009yc}:
\begin{eqnarray}
L_{K} &=& K- \frac{1}{2} S^2 R - \frac{3}{2} S^4 + (\mathrm{fermionic}\ \mathrm{terms})\,,\nonumber \\ L_{R^2} &=&R^2+16S\Box S+12S^2 R +36S^4 +
(\mathrm{fermionic}\ \mathrm{terms})\,.
\end{eqnarray}
If these were the only two invariants then  $L_{\rm Ric}$ would have to be a linear combination of $L_{K}$ and $L_{R^2}$, but  this is not the case!  In
particular, the $RS^2$ terms do not fit. This means that there must exist a third invariant containing $RS^2$ but no curvature-squared terms.  To
construct this invariant we need a globally supersymmetric invariant not containing  a quartic term in the compensating scalar $\phi$. Starting from a
superfield $\Phi= \phi + \theta^\alpha \lambda_\alpha + \theta^2\, S$, one finds that there are two independent superspace actions of this type: \be
{ I}^{\rm rigid}_1 = \int d^3x\, d^2\theta\, (D^2\Phi)^3\, \Phi\,, \qquad { I}^{\rm rigid}_2 =  \int d^3x\, d^2\theta\, (D^2\Phi)^2\, D^\alpha\Phi D_\alpha\Phi\,
. \ee These yield the  component Lagrangians
\begin{eqnarray}
 L^{\rm rigid}_1 & = & S^4 + 48 S^2 \phi \Box \phi - 12 S^2 \bar{\lambda} \partial \hskip -.25truecm \slash
 \lambda - 48 S \phi (\partial_\mu \bar{\lambda}) \gamma^\mu \gamma^\nu (\partial_\nu \lambda) + \cdots \,, \nonumber \\
 L^{\rm rigid}_2 & = & S^4 - 16 S^2 (\partial \phi)^2 - 12 S^2 \bar{\lambda} \partial \hskip -.25truecm \slash \lambda - 32 S \Box \phi \bar{\lambda}
\lambda - 16 (\partial S \cdot \partial \phi) \bar{\lambda} \lambda \nonumber \\ & & + \ 32 S \partial_\mu \phi \bar{\lambda} \gamma^{\mu \nu}
\partial_\nu \lambda + \cdots \,,
\end{eqnarray}
where the dots indicate terms quartic in fermions. The next step consists  in constructing a conformally invariant Lagrangian out of $ L^{\rm rigid}_1$ and
$ L^{\rm rigid}_2$.  It turns out that it is not possible to make them conformally invariant separately; only the combination
\begin{eqnarray}
 L^{\rm rigid}_1 +  9L^{\rm rigid}_2 = 10 S^4 + 48 S^2 \phi \Box \phi - 144 S^2 (\partial \phi)^2 + (\mathrm{fermionic}\ \mathrm{terms})
\end{eqnarray}
can  be made conformally invariant. This follows from the observation that
\begin{equation}
\delta_{K} \left( S^2 \phi \Box^C \phi - 3 S^2 (D \phi)^2 + \frac{1}{16} R S^2 \phi^2\right) = 0 \,.
\end{equation}
The combination $ L^{\rm rigid}_1 + 9 L^{\rm rigid}_2$ can thus be made conformally invariant by taking the following weights:
\begin{equation}
w_\phi = -\frac{1}{4} \,, \quad w_\lambda = w_\phi + \frac{1}{2} = \frac{1}{4} \,, \quad w_S = w_\phi + 1 = \frac{3}{4} \,,
\end{equation}
by turning all derivatives into covariant ones and then adding the curvature-dependent term $3 R S^2 \phi^2$. Upon using the gauge fixing condition
$\phi_0 = 1$, one ends up with the following Lagrangian:
\begin{equation} \label{Sfourth}
L_{S^4} = S^4 + \frac{3}{10} R S^2 + (\mathrm{fermionic}\ \mathrm{terms})\,,
\end{equation}
which was not considered in \cite{Andringa:2009yc}.

The new $S^4$ invariant presented above can be generalized by noting that the following component Lagrangians are also invariant
under rigid supersymmetry:
\begin{eqnarray}
 L^{(n)}_1 & = & S^n + 16 (n-1) S^{n-2} \phi \Box \phi - 4(n-1) S^{n-2} \bar{\lambda} \gamma^\mu \partial_\mu \lambda \nonumber \\ & & -
8(n-1)(n-2) S^{n-3} \phi (\partial_\mu \bar{\lambda}) \gamma^\mu \gamma^\nu (\partial_\nu \lambda) + \cdots \,, \nonumber \\ L^{(n)}_2 & = &
S^n - 16 S^{n-2} (\partial \phi)^2 - 4(n-1) S^{n-2} \bar{\lambda} \gamma^\mu \partial_\mu \lambda - 16(n-2) S^{n-3} \Box \phi \bar{\lambda} \lambda
\nonumber \\ & &- 8(n-2)(n-3) S^{n-4}(\partial S \cdot \partial \phi) \bar{\lambda} \lambda
\nonumber \\ & & + 16(n-2) S^{n-3} \partial_\mu \phi \bar{\lambda}
\gamma^{\mu \nu} \partial_\nu \lambda + \cdots \,.
\end{eqnarray}
Again, only one linear combination of $ L^{(n)}_1$ and $ L^{(n)}_2$ can be made conformally invariant. This conformal combination leads to
the following generalization of (\ref{Sfourth}):
\begin{equation}
L_{S^n}= S^n + \frac{n-1}{6 n - 14} R S^{n-2} + (\mathrm{fermionic}\ \mathrm{terms})\,.
\end{equation}
Choosing $n=1$ we recover the supergravity cosmological term \be L_{S} \equiv L_C = S + (\mathrm{fermionic}\ \mathrm{terms})\, . \ee Choosing $n=2$ we
recover the standard EH terms \be L_{S^2} \equiv -\frac{1}{2} L_{EH} \, , \ee where $L_{EH}$ is given in (\ref{Lpoinc}).  Choosing $n=3$ we arrive at
a new invariant with Lagrangian \be L_{S^3} = S^3 + \frac{1}{2}RS + (\mathrm{fermionic}\ \mathrm{terms})\, . \ee Finally, we recover $L_{S^4}$ of
(\ref{Sfourth}) by choosing $n=4$.


\section{The general `curvature-squared' model}\label{sec:genmod}
\setcounter{equation}{0}


We have now shown that there exist three locally supersymmetric actions with Lagrangians that have the same dimension as $R^2$. The three Lagrangians are
\begin{eqnarray} \label{newlagr}
L_K &=& K- \frac{1}{2} S^2 R - \frac{3}{2} S^4 + (\mathrm{fermionic}\ \mathrm{terms})\,, \nonumber \\
L_{R^2} &=&R^2+16S\Box S+12S^2 R +36S^4 + (\mathrm{fermionic}\ \mathrm{terms})\,,\nonumber\\
L_{S^4} & =& S^4 + \frac{3}{10} R S^2 + (\mathrm{fermionic}\
\mathrm{terms})\,.
\end{eqnarray}
We also found a fourth Lagrangian $L_{\rm Ric}$ of the same dimension but
\begin{equation}
L_{\rm Ric} \equiv L_K + \frac{1}{64} L_{R^2} + \frac{15}{16} L_{S^4} \, .
\end{equation}
In fact, all Lagrangians at this dimension are linear combinations  of $L_K$, $L_{R^2}$ and $L_{S^4}$.  Similarly, at one lower dimension we will have a linear combination of the scalar density $\sqrt{-\det g}\,  L_{S^3}$ and the supersymmetric extension
$\mathcal{L}_{top}$ of the Lorentz-Chern-Simons Lagrangian density ${\cal L}_{LCS}$.

Introducing the gravitational coupling constant $\kappa$, and the notation $e= \sqrt{-\det g}$ for the volume density, we may
now write the action for the most general 3D supergravity with no terms of  dimension higher than $R^2$ as
\begin{eqnarray}\label{genI}
I[g,S] &=& \frac{1}{\kappa^2} \int \, d^3 x \, \left\{ e \left[ M L_C + \sigma L_{EH} + \frac{1}{m^2}L_K + \frac{1}{8 \tilde{m}^2}L_{R^2} +
\frac{1}{{\check m}^ 2} L_{S^4} + \frac{1}{\check \mu} L_{S^3} \right] \right. \nonumber\\
 &&\left.  \qquad \qquad \qquad  +\  \frac{1}{\mu}\mathcal{L}_{LCS} \right\}\, ,
\label{master}
\end{eqnarray}
where  $(M, m, \tilde{m}, \check m)$ are mass parameters, as are $(\mu,\check \mu)$ although the action depends only on the dimensionless combinations
$(\kappa^2\mu,\kappa^2\check \mu)$, and
\be
\mathcal{L}_{LCS} = \frac12 
\varepsilon^{\lambda\mu\nu}\Gamma^\rho_{\lambda\sigma} \left[\partial_\mu\Gamma^\sigma_{\rho\nu} +\frac23\Gamma_{\mu\tau}^\sigma\Gamma_{\nu\rho}^\tau\right]\ .
\ee
The bosonic Lagrangian density is
\begin{eqnarray}\label{genbos}
{\cal L}_{bos} &=&  e \left\{ M S + \sigma \left(R-2S^2\right)  + \frac{1}{m^2}\left(K - \frac{1}{2} RS^2 - \frac{3}{2} S^4\right) +
\frac{1}{\check m^2} \left(S^4 + \frac{3}{10} RS^2\right) \right.\nonumber\\
&&\left. -\   \frac{2}{\tilde{m}^2}\left[ \left(\partial S\right)^2 -
\frac{9}{4} \left(S^2+ \frac{1}{6}R\right)^2\right] + \frac{1}{\check \mu} \left(S^3 +\frac{1}{2} RS\right) \right\} +
\frac{1}{\mu} {\cal L}_{LCS} \, .
\end{eqnarray}
This has six  independent mass parameters $(M, m, \check m, \tilde m, \check\mu,\mu)$, not counting the overall gravitational coupling constant $\kappa$, and one dimensionless constant $\sigma$. In all, there are therefore seven dimensionless parameters. We recall that
we allow $m^2$ to be negative as well as positive, and we will similarly allow $\tilde
m^2$ and $\check m^2$ to take either sign.


\subsection{Some notation}


Before proceeding, we gather together here some useful definitions. First we recall the definition of $\hat m^2$ from \cite{Andringa:2009yc}:
\be
\frac{1}{\hat m^2} = \frac{1}{m^2} - \frac{3}{\tilde m^2}\, .
\ee
Three new definitions are
\bea
 \frac{1}{(\hat m^\prime){}^2} &=&  \frac{1}{\hat m^2}  - \frac{2}{3\check m^2}\, , \nonumber \\
 \frac{1}{(\hat m^\prime{}^\prime){}^2} &=& \frac{1}{\hat m^2}  - \frac{3}{5\check m^2} \, , \nonumber \\
 \frac{1}{(\hat m^\prime{}^\prime{}^\prime){}^2} &=& \frac{1}{\hat m^2} - \frac{27}{40 \check m^2}\, .
 \eea
In the case that $\tilde m^2=\infty$, we drop the hats; for example
\be\label{massparams2}
\frac{1}{(m^\prime){}^2} = \frac{1}{m^2}  - \frac{2}{3\check m^2}\, , \qquad
\frac{1}{(m^\prime{}^\prime){}^2} = \frac{1}{m^2}  - \frac{3}{5\check m^2}\, .
\ee


\subsection{Field equations }


We now turn to the field equations of the general model with Lagrangian density (\ref{genbos}). The $S$ field equation is
\be\label{SFE}
 \left(M-4\sigma S -
\frac{SR}{15 \check m^2} \right) + 3\left(S^2+ \frac{1}{6}R\right)\left(\frac{1}{\check \mu} - \frac{2S}{(\hat m^\prime){}^2}\right) =
-\frac{4}{\tilde m^2} D^2 S \, .
\ee
The metric field equation may be written as
\begin{eqnarray}\label{metricEq}
0 &=& \left(- \frac{1}{2} MS + \sigma S^2 - \frac{S^3}{2\check \mu} +\frac{3S^4}{4 (\hat m^\prime){}^2} \right) g_{\mu\nu} + \sigma G_{\mu\nu} +
\frac{1}{\mu} C_{\mu\nu}  + \frac{1}{2m^2} K_{\mu\nu} + \frac{1}{2\tilde m^2} L_{\mu\nu} \nonumber \\ &&-\ \frac{2}{\tilde m^2} \left[\partial_\mu S
\partial_\nu S - \frac{1}{2} g_{\mu\nu} \left(\partial S\right)^2\right] + \frac{1}{2\check \mu}  \left[ G_{\mu\nu}S - \left(D_\mu D_\nu - g_{\mu\nu}
D^2\right)S\right] \nonumber \\ &&  \ - \ \frac{1}{2(\hat m^\prime{}^\prime){}^2}  \left[ G_{\mu\nu}S^2 - \left(D_\mu D_\nu - g_{\mu\nu}
D^2\right)S^2\right] \, ,
\end{eqnarray}
where (as in \cite{Andringa:2009yc})
\begin{eqnarray}
e\, C_{\mu\nu} &=& \varepsilon_\mu{}^{\tau\rho} D_\tau S_{\rho\nu}\, , \qquad S_{\mu\nu}= R_{\mu\nu}- \frac{1}{4}g_{\mu\nu}R\, ,\\ K_{\mu\nu} &=& 2D^2
R_{\mu\nu} - \frac{1}{2} D_\mu D_\nu R - \frac{1}{2} g_{\mu\nu} D^2R  - \frac{13}{8} g_{\mu\nu} R^2 \nonumber\\ &&+\  \frac{9}{2}R R_{\mu\nu}  -  8
R_\mu{}^\lambda R_{\lambda\nu} + 3 g_{\mu\nu} \left(R^{\rho\sigma}R_{\rho\sigma} \right)\;, \\ L_{\mu\nu} &=& - \frac{1}{2} D_\mu D_\nu R +
\frac{1}{2} g_{\mu\nu} D^2R - \frac{1}{8} g_{\mu\nu} R^2 + \frac{1}{2} RR_{\mu\nu}\, .
\end{eqnarray}
The trace of the metric field equation can be written as
\begin{equation}
\begin{split}
\label{tracemetricgen}
& \left(M-4\sigma S - \frac{1}{15\check m^2} SR\right)S + \left(S^2+ \frac{1}{6}R\right)\left(2\sigma + \frac{S}{\check \mu} + \frac{R}{12\hat
m^2}  - \frac{3S^2}{2(\hat m^\prime){}^2} \right)
\\
&- \  \frac{1}{3m^2}\left(K+ \frac{1}{24}R^2\right)  =  \frac{1}{3\tilde m^2} \left[
2\left(\partial S\right)^2  + D^2R\right] + \frac{2}{3\check \mu} D^2 S
-  \frac{2}{3(\hat m^\prime{}^\prime){}^2} D^2S^2 \, .
\end{split}
\end{equation}


\subsection{Maximally symmetric vacua}


The field equations simplify considerably for maximally-symmetric vacua, which are characterized by the cosmological constant $\Lambda$. The $S$ equation simplifies to
\be\label{vacuumSeq}
\left(M-4\sigma S - \frac{2}{5\check m^2} S\Lambda\right)
+3\left(S^2+\Lambda\right)\left(\frac{1}{\check \mu} - \frac{2S}{(\hat m^\prime){}^2}\right)=0\,.
\ee
{}For maximally symmetric spacetimes, the metric equation is implied by its trace. Using
the fact that \be R= 6\Lambda\, , \qquad K = - \frac{3}{2}\Lambda^2 \, , \ee for maximally symmetric metrics, the trace of the metric equation can be
seen to reduce to
\be\label{tracemetric}
\left(M-4\sigma S - \frac{2}{5\check m^2} S\Lambda\right)S + \left(S^2+\Lambda\right)\left(2\sigma +
\frac{\Lambda}{2\hat m^2} + \frac{S}{\check\mu} - \frac{3S^2}{2(\hat m^\prime){}^2} \right) =0\, .
\ee
Combining this with the $S$ equation, we
deduce that
\be
\left(S^2+\Lambda\right) \left[S^2 -\frac{4(\hat m^\prime){}^2}{9\check\mu} S + \frac{(\hat m^\prime){}^2}{9}\left(4\sigma +
\frac{\Lambda}{\hat m^2}\right) \right] =0\,. \ee There are therefore two classes of maximally symmetric vacua, as found for the less general model
of \cite{Andringa:2009yc} but the present analysis is slightly simpler and better adapted to the more general case now under consideration.  We
consider these two classes in turn.

\begin{itemize}
\item Supersymmetric vacua with \be S^2=-\Lambda \ge0\,. \ee In this case both $S$ and metric equation are solved when $S$ solves the cubic equation
    \be\label{susycubic} M- 4\sigma S + \frac{2}{5 \check m^2} S^3=0\, .
    \ee
    Using the fact that $S^2=-\Lambda$, we can rewrite this cubic
    equation as
    \be M= \left(4\sigma + \frac{2\Lambda}{5\check m^2} \right)S\, .
    \ee
    Squaring both sides we then deduce that
    \be\label{cubfun}
    \Lambda\left(\sigma+ \frac{\Lambda}{10 \check m^2} \right)^2 + \frac{1}{16} M^2=0\,.
    \ee
    This is a cubic function of $\Lambda$ that  can be plotted as
    a curve in the $(\Lambda,M^2)$ plane. In the limit that $\check m^2\to\infty$ this curve reduces to the straight line of
    \cite{Andringa:2009yc} representing supersymmetric vacua.
\item The remaining maximally symmetric vacua are generically non-supersymmetric, and correspond to solutions of the quadratic equation
    \be\label{nonsusy}
    S^2 -\frac{4(\hat m^\prime){}^2}{9\check\mu} S + \frac{(\hat m^\prime){}^2}{9}\left(4\sigma + \frac{\Lambda}{\hat
    m^2}\right)  =0\,.
    \ee
    Using this in (\ref{vacuumSeq}),  we deduce that
    \be
    M - \frac{4(\hat m^\prime){}^2}{27\check\mu} \left(\sigma -
    \frac{20\Lambda}{(\hat m^\prime{}^\prime{}^\prime){}^2}\right) = \frac{4S}{3}\left(\sigma + \frac{4\Lambda}{(\hat
    m^\prime{}^\prime{}^\prime){}^2} - \frac{(\hat m^\prime){}^2}{9\check \mu^2} \right)\, ,
    \ee
    where $S$ is a solution to (\ref{nonsusy}).
       In the limit that  $|\check \mu|  \to \infty$, we have the following cubic equation for $\Lambda$ in terms of $M^2$:
       \be
       \hat m^2 \left(\hat m^\prime{}^\prime{}^\prime\right){}^4 \left(\frac{9M}{16}\right)^2 = - \left(\hat
    m^\prime\right){}^2 \left(\Lambda + 4 \hat m^2\sigma\right)\left[\Lambda +
    \frac{1}{4}\left(\hat m^\prime{}^\prime{}^\prime\right){}^2 \sigma\right]^2\, .
    \ee
    As expected,  the sign of $M$ is relevant only when $\check \mu$ is finite because otherwise the
    field redefinition $S\to -S$ flips the sign of $M$ without causing any other change.
    In the further limit that $\check m^2\to \infty$, the cubic reduces to the cubic found in \cite{Andringa:2009yc} and
    plotted there in the  $(\Lambda, M^2)$ plane.

\end{itemize}


\subsection{Review of supersymmetry-preservation  conditions}
\label{subsec:susyrev}


The necessary and sufficient conditions for any bosonic field configuration of 3D supergravity to be supersymmetric were found in
\cite{Andringa:2009yc}. We shall review the result here as we will want to know whether the solutions of the field equations that we consider are
supersymmetric solutions.  A useful necessary condition for supersymmetry is that \be\label{necessary} 16 \left(\partial S\right)^2 =
\left(R+6S^2\right)^2\,. \ee When $S$ is constant this implies that $R+6S^2=0$, and this reduces to the condition (\ref{SLam}) for maximally symmetric
vacua,  defined by the condition (\ref{maxsym}). In this case, one can show (by constructing the Killing spinors) that maximally symmetric vacua
satisfying (\ref{SLam}) are also maximally supersymmetric.

More generally, a bosonic configuration of 3D supergravity is supersymmetric if the metric and scalar field $S$  take the form
\be\label{susymetric}
ds^2 = dx^2 + 2f(u,x) \, du dv + h(u,x) du^2\ , \qquad S= - \partial_x \log \sqrt{f}\ , \ee where the functions $f$ and $h$ are arbitrary, except that $f$
is nowhere vanishing. This implies that
\be
\partial_x S = \frac14( R+ 6S^2) \, ,
\ee
which is obviously compatible with (\ref{necessary}) but is a stronger condition.

All cases that we will consider here have {\it constant} $S$; in this case the configuration (\ref{susymetric}) can be put into the form
\be
ds^2=
dx^2 + 2e^{\mp 2x/\ell} dudv + h(u,x)du^2 \, , \qquad S= \pm \ell^{-1}\, ,
\ee
for constant $\ell$ (with dimensions of inverse mass).   Introducing the new coordinate
\be
r= e^{\mp x/\ell}\, ,
\ee
we see that the supersymmetric configurations for constant $S$ can be put into the  pp-wave form
\be\label{ppwavemetric}
 ds^2 = \ell^2 \frac{dr^2}{r^2} + 2r^2 dudv + h(u,r) du^2 \, , \qquad S= \pm \ell^{-1}\, .
\ee
When $h=0$ we have an adS spacetime  with adS radius $\ell$.


\subsection{The pp-wave solution revisited} \label{secppwave}


We know from \cite{Andringa:2009yc} that there are supersymmetric pp-wave configurations, of the type first discussed in \cite{Deser:2004wd},
that solve the equations of motion of the curvature-squared supergravity model constructed there. We now investigate this issue in the
 context of the more general model.
To this end, we first rewrite the metric of (\ref{ppwavemetric}) as
\be
ds^2= 2e^+e^- + e^*e^*\, , \qquad S= \pm \ell^{-1}\, ,
\ee
where
\be\label{basis}
e^+ =r dv
+\frac{h(u,r)}{2r} du\ ,\qquad e^- = r du\ ,\qquad e^*=\frac{\ell}{r} dr\, .
\ee
The non-vanishing components of the Ricci  and Cotton tensors  are
\bea R_{+-} &=&R_{**}\ = \ -2\ell^{-2}\ , \qquad  R_{--} =  -\frac{\ell^{-2}}{2r^2} \left(r^2\partial_r^2 -r\partial_r\right)h\
, \nonumber \\
C_{--} &=& \ell^{-1}(r\partial_r +1)R_{--}\ . \label{e2}
\eea
The Ricci scalar is then given by $R=-6/\ell^2$.

Using these results, we find that the $S$ field equation (\ref{SFE})
reduces to
\be
M  \mp 4\sigma\ell^{-1}  \pm \frac{2\ell^{-3}}{5\check m^2}=0\ .
\ee
We also find that all components of the metric equation (\ref{metricEq})  are satisfied trivially except the $--$ component, which gives
\be
\label{einstein}
\left[
\frac{1}{m^2}r^2\partial_{r}^2+\left(\frac{3}{m^2}+ \frac{\ell}{\mu}\right)r\partial_{r} + \ell^2 \left(\hat\sigma +\frac{1}{\mu\ell}\right)\right]
\left[\partial_r^2 - \frac{1}{r} \partial_r\right] h =0\,  ,
\ee
where
\be\label{sihat}
\hat\sigma =
\sigma \pm \frac{\ell^{-1}}{2\check \mu} +\frac{3\ell^{-2}}{10\check m^2}\, .
\ee

Trying a solution of the form  $h\propto  r^n$, we find that it solves the fourth-order ODE as long as the power $n$ satisfies the quartic characteristic equation
\be
n(n-2) \left( \frac1{m^2} n(n-2) + \frac{\ell}{\mu} (n-1) + \ell^2\hat\sigma\right)=0\ ,
\label{ce}
\ee
which has roots $0,2,n_+,n_-$, where
\bea
n_\pm = 1-\frac{\ell m^2}{2\mu}\pm
\sqrt{1+\frac{m^4\ell^2}{4\mu^2}- m^2\ell^2 \hat\sigma }\,  .
\label{npm}
\eea
Thus, the generic supersymmetric pp-wave solution has
\bea
h(u,r) = h_+(u) \ell^{2-n_+} r^{n_+} + h_-(u) \ell^{2-n_-}r^{n_-}  + r^2f_{2}(u)+ \ell^2 f_{3}(u)\, , \label{pp}
\eea
where $h_\pm, f_2, f_3$ are arbitrary dimensionless functions of $u$. One can arrange for $f_2$ and $f_3$ to vanish by local coordinate transformations, so
the solution is essentially determined by the two dimensionless functions $h_\pm(u)$.

The solution (\ref{pp}) assumes that the four roots $0$, $2$, $n_+$ and $n_-$ are all different. Several critical points can be identified, where some of these roots become degenerate. We can distinguish the following cases:

\begin{itemize}
\item $n_+=n_-$, $n_\pm \neq 0,2$

In this case the characteristic equation has a doubly degenerate root; this arises for
\be
m^2=  2\mu^2 \left( \hat\sigma \pm
\sqrt{\hat\sigma^2 -\frac1{\ell^2\mu^2}}\right) \equiv m^2_\pm \,  ,
\label{slog}
\ee
in which case the generic solution  (after setting $f_2=f_3=0$) is
\be
h(r,u) =   \ell^{2-k_\pm} r^{k_\pm} \left[h_1(u) \log \left(r/\ell\right) + h_2(u) \right]\,  , \label{log}
\ee
where
\be
k_\pm  \equiv   1-(\ell m^2_\pm/ 2\mu) =  1-\ell\mu\hat\sigma \mp \sqrt{\ell^2\mu^2\hat\sigma^2-1}\, ,
\ee
and $h_1(u), h_2(u)$ are arbitrary dimensionless functions of $u$.

\item $n_- = 0$ or $n_- = 2$, $n_+ \neq 0,2$

This case occurs when $\ell \mu \hat{\sigma} = +1$ (for $n_-=0$) or $\ell \mu \hat{\sigma} = -1$ (for $n_- = 2$). In case the root 0 becomes doubly degenerate, the generic solution (with $f_2 = f_3 = 0$) is
\be
h(r,u) = \ell^{2-k_1} r^{k_1} h_1(u) + \ell^2 h_2(u) \log(r/\ell) \,,
\label{log1}
\ee
where
\be
k_1 = 2 - \frac{\ell m^2}{\mu} \,,
\ee
and $h_1(u), h_2(u)$ are arbitrary dimensionless functions of $u$. For $n_- = 2$, the generic solution is given by
\be
h(r,u) = \ell^{2-k_2} r^{k_2} h_1(u) + r^2 \log(r/\ell)h_2(u) \,,
\label{log2}
\ee
where
\be
k_2 = - \frac{\ell m^2}{\mu} \,.
\ee

\item $n_+ = 0$ or $n_+ = 2$, $n_- \neq 0,2$

This case is analogous to the previous one, with $n_-$ and $n_+$ interchanged. It thus occurs when $\ell \mu \hat{\sigma} = +1$ (for $n_+=0$) or $\ell \mu \hat{\sigma} = -1$ (for $n_+ = 2$). The generic solutions are given by (\ref{log1}) (for $n_+ = 0$) and (\ref{log2}) (for $n_+ = 2$).

\item We can also consider the case for which the roots $n=0$ and $n=2$ become triply degenerate. The conditions $n_+ = n_- = 0$ occur for $\ell\mu\hat\sigma=1$ and $\ell m^2 = 2 \mu$, while $n_+ = n_- = 2$ is obtained by taking $\ell\mu\hat\sigma=-1$ and $\ell m^2 = -2 \mu$. At these critical points, the pp-wave solutions disappear and become diffeomorphic to ${\rm adS}_3$. New doubly logarithmic solutions arise given by
\bea
&& \ell\mu\hat\sigma=+1:\qquad h(r,u)=\ell^2 \log \left(r/\ell\right)  \left[ h_1(u) \log \left(r/\ell\right) + h_2(u)\right]\, , \nn
&&
\ell\mu\hat\sigma=-1:\qquad h(r,u)= r^2 \log \left(r/\ell\right)   \left[ h_1(u) \log \left(r/\ell\right)
+ h_2(u)\right]\, , \label{doublelog}
\eea
where, again,  $h_1(u), h_2(u)$ are arbitrary dimensionless functions of $u$.
\end{itemize}

All of the pp-wave solutions presented above reduce to those found
in \cite{Andringa:2009yc} in the limit $\check\mu\rightarrow \infty$
and $\check m^2 \rightarrow \infty$, and for $h\ne 0$ they all
preserve half the supersymmetry of the ${\rm adS}_3$ vacuum with
$\Lambda= -1/\ell^2$; in the conventions of \cite{Andringa:2009yc}
the Killing spinor is
\be \epsilon_{Kill.} = \sqrt{\frac{r}{\ell}}\left(
  \begin{array}{c}
    \psi_0 \\
    0 \\
  \end{array}
\right)\, ,
\ee
where $\psi_0$ is an arbitrary constant. For $h=0$ the solution degenerates to the supersymmetric ${\rm adS}_3$ vacuum, which preserves both supersymmetries; the generic Killing spinor now takes the form
 \be \epsilon_{Kill.} =  \sqrt{\frac{r}{\ell}} \left(
  \begin{array}{c}
     \psi_0 + {\sqrt 2}  v\chi_0 \\
 \ell\chi_0/r
  \end{array}
\right)\, , \ee
for arbitrary constants  $\psi_0$ and $\chi_0$.


\section{Models with auxiliary $S$}
\label{sec:superGMG}
\setcounter{equation}{0}


In this section we will study special cases of the model defined by \eq{genbos} for which
\be\label{nodS2}
\tilde m^2 =\infty\, .
\ee
This defines a six-parameter subclass of models, all  with the feature that the equation for $S$ is algebraic, in fact a cubic equation.
However, the coefficients are not necessarily constant and this will generically lead to a propagating scalar mode. This can be avoided by imposing additional conditions on the parameters that define the following classes of models:
\bea
{\rm Super-GMG}: \qquad
\ \  &&  {\tilde m}^2=\infty\, , \quad \check\mu^2 =\infty\, \quad   (m^\prime{}^\prime){}^2 =\infty \label{defs}\\
{\rm Super-NMG}: \qquad
\ \   && {\tilde m}^2 =\infty\, , \quad  \check \mu^2 =\infty\, , \quad
(m^\prime{}^\prime){}^2 =\infty\, , \quad |\mu|=\infty \nonumber
\eea
Note that
\be
(m^\prime{}^\prime){}^2 =\infty \quad \Leftrightarrow \quad  \check m^2 = \frac{3}{5} m^2\, .
\ee
We shall see that there are other ``generalized'' cases, with finite  $\check{\mu}$,  in which a propagating scalar can be avoided,
but these arise as a consequence of a relation between the parameters of the model and the vacuum value of $S$;  see eq.~(\ref{balance})  below.


\subsection{Super-GMG}\label{subsec:superGMG}

We begin with the super-GMG model. In this case the  Lagrangian density (\ref{genbos}) simplifies to
 \be\label{superGMGLag}
 {\cal L} = e\left\{ \left(MS -2\sigma
S^2 + \frac{1}{6m^2}S^4\right) + \sigma R + \frac{1}{m^2}K\right\} +
\frac{1}{\mu} {\cal L}_{LCS}\, , \ee which contains the four
independent parameters $M\,,\sigma\,, m$ and $\mu$. The $S$ equation
of motion is the cubic equation \be\label{aux} M- 4\sigma S +
\frac{2S^3}{3m^2} =0\, . \ee The special feature of super-GMG is
that the coefficients of this cubic equation are constants, which
means that  $S$ is constant. There is always at least one solution,
and it is unique  when \be \label{cases} 9M^2 >  128\, m^2\sigma^3\,
. \ee This is satisfied automatically when $m^2\sigma<0$.

Given a solution $S=\bar S$ of (\ref{aux}), back-substitution into
the Lagrangian density yields\footnote{As the equation for $S$ is
cubic rather than quadratic, this back-substitution is not
equivalent to Gaussian integration over $S$ in the path integral.
However, substitution into the field equations rather than the
action $I[g,S]$ yields equations that are equivalent to those  found
from the action $I[g,\bar S]$, so the back-substitution is still
justified classically.} \be\label{GMGlag} {\cal L}_{GMG} = e\left\{
-2\lambda m^2  + \sigma R + \frac{1}{m^2}K\right\} + \frac{1}{\mu}
{\cal L}_{LCS}\, , \ee where $\lambda$ is related to $\bar S$ via
the quartic equation
 \be\label{lameq} 4m^4\lambda = \bar S^4 -4m^2\sigma \bar S^2 \, .
 \ee
 This is just the cosmological GMG Lagrangian density of \cite{Bergshoeff:2009hq}, hence the terminology ``super-GMG'' for the model with
 bosonic Lagrangian density (\ref{superGMGLag}). The special case in which $|\mu|=\infty$ is then ``super-NMG''.


\subsubsection{Field equations and vacua}


The metric equation of the general model simplifies enormously for super-GMG:
\be\label{metricSGMG}
 \left(- \frac{1}{2} MS + \sigma S^2 - \frac{1}{12m^2} S^4\right) g_{\mu\nu} + \sigma G_{\mu\nu} + \frac{1}{\mu} C_{\mu\nu}  +
 \frac{1}{2m^2} K_{\mu\nu} =0\, .
\ee
The trace of this equation can be written as
\be
 \left(M-4\sigma S + \frac{2S^3}{3m^2} \right)S +  \left(S^2+ \frac{1}{6}R\right)\left(2\sigma + \frac{R}{12m^2} - \frac{S^2}{2m^2}\right)
 = \left(K+ \frac{1}{24}R^2\right)\, .
 \ee
Remarkably, the first-parenthesis terms vanish on using the $S$ field equation (\ref{aux}). Given that $S=\bar S$ solves that cubic equation, we see that the trace of the metric equation further simplifies to
\be
\label{tracemetricSGMG} \left(\bar S^2+ \frac{1}{6}R\right)\left(2\sigma +
\frac{R}{12m^2} - \frac{\bar S^2}{2m^2}\right)
 = \left(K+ \frac{1}{24}R^2\right)\, .
\ee
For maximally symmetric vacua, for which
\be
\label{KrelR} K+ \frac{1}{24}R^2 =0\, ,
\ee
this equation reduces to
\be
\label{fork} \left(\bar S^2 + \Lambda\right)\left(4m^2\sigma + \Lambda- \bar
S^2\right) =0\, ,
\ee
which also follows from a comparison of (\ref{lameq}) with (\ref{lamLam1}). There are therefore two classes of  vacua of super-GMG:
\begin{itemize}

\item Supersymmetric vacua with $\bar S^2 =-\Lambda$. In this case \be 9m^4M^2 = -4\Lambda\left(\Lambda+6m^2\sigma\right)^2\, , \ee with
    $\Lambda < 0$, so these vacua are either Minkowski or adS.

\item Non-supersymmetric vacua with  $\bar S^2= 4m^2\sigma + \Lambda \ne -\Lambda$. In this case \be 9m^4M^2  = 4\left(\Lambda +
    4m^2\sigma\right)\left(\Lambda-2m^2\sigma\right)^2\, , \ee with $\Lambda>-4m^2\sigma$; for $m^2\sigma<0$ this implies that all
    non-supersymmetric vacua are dS, but there are also non-supersymmetric adS vacua (with $\lambda<0$) when $m^2\sigma>0$.

\end{itemize}
A consequence of the restriction on $\Lambda$ in each of these cases is that $\lambda\ge0$ when $m^2\sigma<0$. Thus, not all of the vacua of GMG are vacua of super-GMG; the dS vacua for $\lambda<0$ and $m^2\sigma<0$ are excluded.

As a simple illustration of the fact that there exist supersymmetric adS vacua, consider \be m^2\sigma< 0 \, , \qquad M^2 \ll \left|m^2\sigma\right|\,
. \ee In this case there is a unique solution $\bar S$ of the cubic equation (\ref{aux}), and it  takes the form \be \bar S= \frac{M}{4\sigma}\left[1
+ \frac{M^2}{96m^2\sigma^3} +  {\cal O}\left(\frac{M^4}{m^4\sigma^2}\right)\right]\, . \ee The cosmological constant is therefore \be \Lambda= -
\frac{M^2}{16\sigma^2}\left[ 1+ \frac{M^2}{48m^2\sigma^3} +   {\cal O}\left(\frac{M^4}{m^4\sigma^2}\right) \right] \le0 \, . \ee It follows that $\bar
S^2=-\Lambda$ to the approximation at which we are working, whereas $\bar S^2 \ne \Lambda + 4m^2\sigma$ within the same approximation. We thus deduce
that these adS vacua are supersymmetric. The limit $M\to 0$ yields the supersymmetric Minkowski vacuum.

To proceed further, it is convenient to define the two dimensionless parameters \be\label{xy} y= \frac{9M^2}{32 m^2\sigma^3}\, , \qquad x= 1+
\frac{\Lambda}{2m^2\sigma}\, . \ee Note that $y\ge 0$ when $m^2\sigma>0$ and $y\le 0$ when $m^2\sigma<0$, and hence that $m^2\sigma$ may have either
sign when $y=0$.


\begin{figure}[tb]
\includegraphics[width=12truecm]{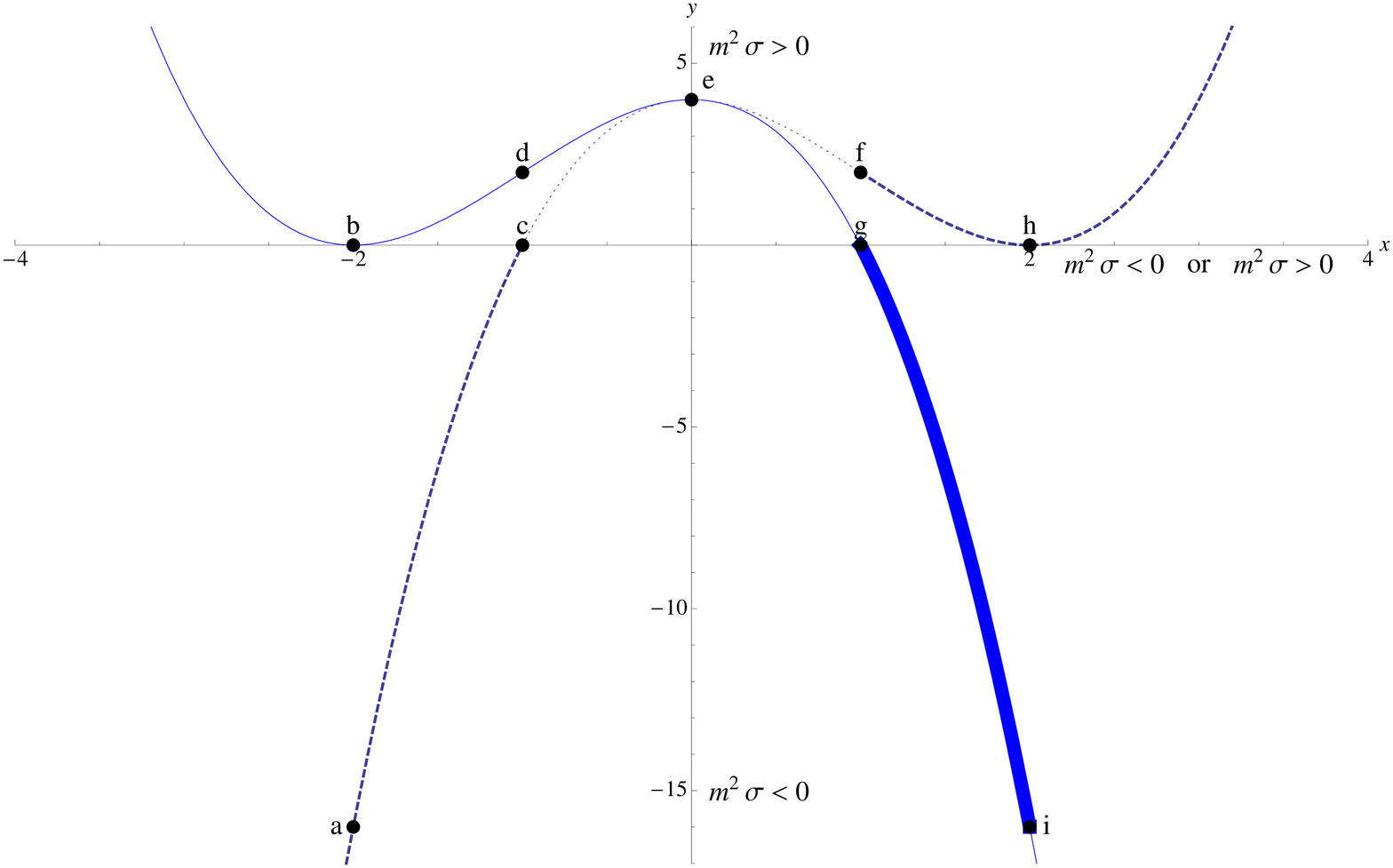} \vspace{-2cm}
\caption{Graphical representation of the maximally-symmetric vacua of super-GMG in the $(x,y)$-plane, with $x$ and $y$ defined in
(\ref{xy}). Supersymmetric vacua correspond to points on the solid curve $y=4-3 x^2 - x^3$; all are adS except for the special point $g$ on this curve, which is Minkowski. The thick part of this curve  corresponds to supersymmetric adS vacua in which NMG is perturbatively unitary, as discussed in section \ref{sec:bulkunitarity}. All other vacua correspond to points on the dashed/dotted curve $y=4-3 x^2 + x^3$.  Those on the (thick) dashed line are dS, while those on the (thin) dotted line are adS. The points $a$ and $h$ are dS vacua, while $b$, $d$, $e$ and $i$ are supersymmetric adS. The point  $f$ is a non-supersymmetric Minkowski vacuum. The point $c$ can be dS or non-supersymmetric adS depending on the sign of $m^2 \sigma$.} \label{figvacua}
\end{figure}

All maximally-symmetric vacua correspond to points in the
$(x,y)$-plane that lie on one of the two cubic curves
\be
y = 4-3x^2 \mp x^3\, ,
\ee
where the upper sign yields the supersymmetric
vacua. Taken together, these two cubic curves yield a figure in the
$(x,y)$ plane, as shown in Fig 1. This figure is  symmetric under
$(x,y)\to (-x,y)$, although this transformation exchanges a
supersymmetric with a non-supersymmetric vacuum, except at the fixed
point $(x,y)=(0,4)$ where the two cubic curves cross. This crossing
point corresponds to a supersymmetric adS vacuum with
$\lambda=-\sigma^2$, as follows from \be\label{lamx} \lambda +
\sigma^2 = \sigma^2 x^2 \, . \ee This is the unique vacuum on the
$y$-axis, from which we deduce that the dS vacuum of cosmological
GMG with $\lambda=-\sigma^2$ and $m^2\sigma<0$ is {\it not} a
solution of super-GMG. As pointed out in \cite{Bergshoeff:2009aq},
the adS vacuum at $\lambda=-\sigma^2$  and $m^2\sigma>0$ has very
special properties; in particular it admits a class of
asymptotically adS black hole solutions, with the extremal black
hole solution interpolating between the adS vacuum and a
Kaluza-Klein solution with ${\rm adS}_2\times S^1$ spacetime (see
also \cite{Oliva:2009ip,Clement:2009gq}).

Let us now consider the possible vacua on each of the two cubic curves separately.  All points on the `supersymmetric'  cubic curve correspond to adS
vacua except, of course, the point at which this curve crosses the $x$-axis; at this point $x=1$, so $\Lambda=0$. This is the supersymmetric Minkowski
vacuum with $M=0$, and $\lambda=0$, although we could consider  this point as representing two vacua since it is valid for either choice of sign of
$m^2\sigma$. There is also a supersymmetric adS vacuum for $m^2\sigma>0$ when $M=0$; this corresponds to the point $(x,y)=(-2,0)$ at which the curve
just touches the $x$-axis. This has $\Lambda=-6m^2\sigma$, and $\lambda=3\sigma^2$.

The analogous analysis for  points on the `non-supersymmetric' cubic curve is a little more complex. Points on this curve with $|x|>1$ correspond to
dS vacua, either with $m^2\sigma>0$ (for $y>0$) or $m^2\sigma<0$ (for $y<0$). The limiting point $(x,y)= (1,2)$  corresponds to a non-supersymmetric
Minkowski vacuum with $m^2\sigma>0$ and  $\lambda=0$. The other limiting point $(x,y)= (-1,0)$ corresponds to a dS vacuum with $m^2\sigma<0$ and
$\lambda=0$ if it is approached from the $y<0$ side. However, it  can also be approached from the $y>0$ side,  in which case it corresponds to an adS
vacuum with  $m^2\sigma>0$ and $\lambda=0$. Elsewhere on this cubic curve, i.e. for $y>0$ and $x<1$, points on the curve correspond to adS vacua that
are not supersymmetric except at the crossing point $(x,y)=(0,4)$.

To make contact with the analysis in \cite{Bergshoeff:2009aq} of the maximally-symmetric vacua of GMG, we first recall that (\ref{lamLam1}) has the
solution \be \Lambda = -2m^2\left[ \sigma \pm \sqrt{\sigma^2+ \lambda}\right]\, , \ee which shows that there are two possible vacua for each $\lambda>
-\sigma^2$.  However, this becomes 4 vacua  for each $\lambda$ if one allows either sign of $m^2\sigma$.   This result  is manifest  from Fig. 1 since
each value of $\lambda>-\sigma^2$ corresponds to two (vertical) lines in the $(x,y)$ plane that are parallel to, but not coincident with, the
$y$-axis, and each of these vertical lines cuts each of the two cubics curves once. Actually, this is not quite right for $\lambda=0$, but let us postpone consideration of this special case, and illustrate  the  generic case with $\lambda=3\sigma^2$, which corresponds to $x= \pm2$. The choice $x=2$ yields a non-supersymmetric dS vacuum at
$(x,y)=(2,0)$ (and hence  $\Lambda = 2m^2\sigma >0$ and $M=0$)  and a supersymmetric adS vacuum  at $(x,y)=(2,-16)$ (and hence $\Lambda= 2m^2\sigma
<0$ and $M\ne0$). As shown in \cite{Bergshoeff:2009aq}, the latter vacuum has very special properties; in particular, linearization about it yields a
quadratic model describing massive particles of spin 1 rather than spin 2.  The other choice $x=-2$ yields a supersymmetric adS vacuum at  $(x,y)=(-2,0)$
(and hence $\Lambda=-6m^2\sigma<0$ and $M=0$) and a dS vacuum at $(x,y)=(-2,-16)$ (and hence $\Lambda=-6m^2\sigma>0$ and $M\ne0$). There is complete
agreement with \cite{Bergshoeff:2009aq} and we now learn that the two adS vacua are supersymmetric in the context of GMG.

The $\lambda=0$ case, which corresponds to $|x|=1$,  is special because the point $(x,y)= (-1,0)$ represents two possible non-supersymmetric vacua,
either dS or adS, depending on the sign of $m^2\sigma$,  as we already observed above, and the same can be said of  the point $(x,y)=(1,0)$ although
both vacua are Minkowski. Taking this into account, we have six vacua for $\lambda=0$. One may ask how this is compatible with our earlier conclusion that each value of $\lambda>-\sigma^2$ corresponds to four distinct vacua,  allowing for either sign of $m^2\sigma$. The answer to this question is that two vacua may be equivalent in the context of GMG but  distinct in the context of super-GMG.  For example, in the GMG context the adS vacuum at $(x,y)=(-1,2)$ would have to be considered  equivalent to the  adS vacuum at $(x,y)= (-1,0)$ because both have the same value of $\Lambda$ and
$\lambda$. But these two vacua have different values of $M^2$ in the super-GMG context; moreover,  one is supersymmetric and the other is not.
Similarly, the Minkowski vacuum at $(x,y)= (1,2)$ is equivalent to the $m^2\sigma>0$ Minkowski vacuum at  $(x,y)= (1,0)$  in the GMG context, but they differ as vacua of super-GMG because they again have different values of $M^2$ and one is supersymmetric and the other not.


\subsubsection{Other solutions}


Let us now turn to solutions of super-GMG that are not maximally symmetric. Of particular interest are solutions that preserve some fraction of the
supersymmetry of a supersymmetric vacuum solution; this fraction is necessarily either $1/2$ or $1$.  Let us begin with the observation that since
$S=\bar S$, a constant, in any solution of super-GMG (in contrast to the general model) all supersymmetric solutions have
\be R= -6 \bar S^2\, .
\ee
Using this to eliminate $R$ from (\ref{KrelR}), we deduce that
\be
K= - \frac{1}{24}R^2 = - \frac{3}{2} \bar S^4\, .
\label{Krel}
\ee
In other words, both $R$ and $K$ must be constants,  such that the vacuum relation (\ref{Krel}) holds. This is a very strong condition that eliminates some otherwise plausible candidate solutions.

For example, for the special case of $\lambda=-1$ and $m^2\sigma>0$,
for which there is a unique adS vacuum, there is also an ${\rm
adS}_2\times S^1$ `Kaluza-Klein' vacuum \cite{Clement:2009gq}. In
this vacuum
\be
R= -4m^2\sigma\, , \qquad K= 2m^4\sigma^2\, .
\ee
Since the relation (\ref{Krel}) does not hold, this vacuum is not supersymmetric. It follows immediately that the static extreme black hole that interpolates between the adS vacuum (at
infinity) and the  `Kaluza-Klein' vacuum (near the horizon) \cite{Bergshoeff:2009aq} is also not supersymmetric.

GMG has extremal BTZ black holes that are supersymmetric solutions of super-GMG. This is because, firstly, the BTZ black holes are isometric to an adS vacuum and hence solutions of super-GMG (because all adS vacua of GMG are solutions) and, secondly, because the analysis of whether global identifications of adS preserve some fraction of supersymmetry is independent of the choice of action. This argument actually applies to the general curvature-squared model, but we concentrate on super-GMG. Are there any other supersymmetric black holes?

To be supersymmetric a black hole must have zero Hawking temperature. This immediately excludes the class of stationary black hole solutions of NMG found in \cite{Clement:2009gq}. It does not exclude the class found in  \cite{Clement:2009ka}, which all have zero Hawking temperature, but we have not attempted to determine whether any of these are supersymmetric; it would be a surprise if they were given the absence of
non-BTZ supersymmetric static black holes.


\subsection{Generalized super-GMG}
\label{subsec:genGMG}


We turn now to the more general models for which $S$ is auxiliary.
Given only the condition (\ref{nodS2}), the bosonic truncation of
the general action (\ref{genI}) is \bea\label{genGMG} I[g,S] &=&
\frac{1}{\kappa^2}\, \int d^3 x\, \left\{ e\left[ \left(MS - 2\sigma
S^2 + \frac{S^3}{\check\mu} - \frac{3S^4}{2(m^\prime)^2} \right) +
\sigma R + \frac{1}{m^2}K  \right. \right.\nonumber \\
&&\left.\left.  \qquad \qquad +\ \frac{1}{2\check\mu}RS - \frac{1}{2(m^\prime{}^\prime)^2}RS^2 \right] +
\frac{1}{\mu} {\cal L}_{LCS}\right\}\, ,
\eea
where $m^\prime$ and $m^\prime{}^\prime$ are as defined in (\ref{massparams2}). The $S$-equation of motion is algebraic:
\be\label{algS}
M - 4\sigma S + \frac{3S^2}{\check \mu} - \frac{6S^3}{(m^\prime)^2} = \left( \frac{S}{(m^\prime{}^\prime)^2} - \frac{1}{2\check \mu}\right)R\, ,
\ee
and it can be solved as a power series in $R$ as long as
\be
0\ne A\equiv 2\sigma - \frac{3\bar S}{\check\mu} + \frac{9\bar S^2}{(m^\prime)^2}\, .
\ee
To see this, we set
\be
S= \bar S + \alpha R + \frac{1}{2}\beta R^2 + {\cal O}\left(R^3\right)\, ,
\ee
where $\bar S$ is a constant solution of the cubic equation
\be\label{cubic-genGMG}
M - 4\sigma \bar S + \frac{3\bar S^2}{\check \mu} - \frac{6\bar S^3}{(m^\prime)^2} = 0\, .
\ee
Substitution into (\ref{algS}) yields
\be
\alpha = \frac{1}{2A} \left(\frac{1}{2\check \mu} -  \frac{S}{(m^\prime{}^\prime)^2}\right)\, , \qquad
\beta =  \frac{\alpha}{A} \left[ 3\alpha \left(\frac{1}{\check\mu} - \frac{6\bar S}{(m^\prime)^2} \right) - \frac{1}{(m^\prime{}^\prime)^2}\right]\, .
\ee
There is no solution of the assumed form if $A=0$; in this case the series must involve fractional powers of $R$.  Assuming $A\ne0$, elimination of $S$  yields  a Lagrangian density of  the form
\be\label{Lag-genGMG}
{\cal L} = e\left[-2\bar\lambda m^2 + \bar\sigma R + \frac{1}{m^2}K\right] + \left(\frac{1}{2\check\mu} - \frac{\bar S}{(m^\prime{}^\prime)^2}\right)^2 \frac{R^2}{4A} + {\cal O}\left(R^3\right) +  \frac{1}{\mu} {\cal L}_{LCS}  \, ,
\ee
where
\be
-2\bar\lambda m^2 = M\bar S - 2\sigma \bar S^2 + \frac{\bar S^3}{\check\mu} - \frac{3\bar S^4}{2(m^\prime)^2} \, , \qquad
\bar\sigma = \sigma + \frac{\bar S}{2}\left(\frac{1}{\check\mu} - \frac{\bar S}{(m^\prime{}^\prime)^2}\right)\, .
\ee

We now have a model that involves, generically,  an additional $R^2$ term as  compared with GMG, as well as  higher powers of $R$. This leads  to a loss of perturbative unitarity in a Minkowski vacuum and we shall see in the following section that the same is true for an  adS vacuum. However,
the additional $R^2$ term in the action is absent in the special case that
\be\label{balance}
\frac{1}{2\check\mu} = \frac{\bar S}{(m^\prime{}^\prime)^2}\, ,
\ee
and it is then obvious from (\ref{algS}) that all higher powers of $R$ are also absent. The Lagrangian density (\ref{Lag-genGMG}) is therefore {\it precisely} of  GMG form in this case, with coefficients
\be
-2\bar\lambda m^2 = M\bar S -2\sigma \bar S^2 - \frac{3\bar S^4}{2(m^\prime)^2} + \frac{2\bar S^4}{(m^\prime{}^\prime)^2}\, , \qquad
\bar\sigma =\sigma + \frac{\bar S^2}{2(m^\prime{}^\prime)^2}\, .
\ee
For the analysis of the following section,  it is convenient to introduce the new dimensionless parameter
\be\label{adef}
a= 2m^2\ell\left(\frac{\bar S}{\left(m^\prime{}^\prime\right)^2} - \frac{1}{\check\mu}\right)
\ee
The condition (\ref{balance}) can then be written more simply as $a=0$. This condition defines what we shall call the  ``generalized super-GMG'' case.
We say ``case''  rather than ``model'' because the condition (\ref{balance}) is not just a relation between the parameters of the general `auxiliary-$S$' model but also involves  $\bar S$.

Observe that one way to achieve $a=0$ is to set $(m^\prime{}^\prime)^2 = \infty$ and  $|\check\mu|=\infty$. We can view this as the special case in which both $a=0$ and $|\check\mu|=\infty$ since these two conditions imply $(m^\prime{}^\prime)^2 = \infty$. What is special about it is that no condition is imposed on
$\bar S$, so we have a relation between the parameters of the general `auxiliary-$S$' model that define a subclass of models. This is precisely the ``super-GMG'' subclass, which therefore arises as the $|\check\mu|=\infty$ subcase of the $a=0$ ``generalized super-GMG'' case. Except for this special subcase,  $\bar S$ is constrained by the relation
\be
\bar S = (m^\prime{}^\prime)^2/2\check \mu\, .
\ee
Consistency with  (\ref{cubic-genGMG}) then requires that
\be
\check \mu M= (m^\prime{}^\prime)^2 \left(2\sigma - \frac{(m^\prime{}^\prime)^4}{20\check\mu^2 \check m^2}\right)\, .
\ee
If the various mass parameters of the model defined by (\ref{genGMG}) satisfy this equation then there exists a (constant) solution $\bar S$ of the equation for $S$ for which $I[g,\bar S]$ is a GMG action. One simple way in which this condition on the parameters can be satisfied is to take $\check m^2=\infty$ and $\check\mu M= 2\sigma m^2$.


\section{Perturbative unitarity of generalized super-NMG} \setcounter{equation}{0}
\label{sec:bulkunitarity}


We now turn to the issue of linearized perturbations about supersymmetric adS vacua. One of our purposes is to make contact with the results of \cite{Bergshoeff:2009aq} on  linearized perturbations of NMG about adS vacua. The auxiliary tensor field method used there was covariant,  off-shell,  and led to complete results that were easy to interpret. Here we show how this method applies to super-NMG, and extend it to deal with the generalized super-NMG case. However, we take as our starting point the generic parity-preserving `auxiliary-$S$' model  for which the Lagrangian density is obtained by taking the $|\mu|\to \infty$ in (\ref{genGMG}):
\be\label{genNMG}
{\cal L} =  e\left[ \left(MS - 2\sigma S^2 + \frac{S^3}{\check\mu} - \frac{3S^4}{2(m^\prime)^2} \right) + \sigma R  + \frac{RS}{2\check\mu} - \frac{RS^2}{2(m^\prime{}^\prime)^2}
+ \frac{1}{m^2}K\right] \, .
\ee
As explained in the previous section, elimination of $S$ leads generically to an infinite  series in powers of $R$. As each term could contribute to the quadratic approximation in an expansion about an adS vacuum, it is simpler to retain $S$ as an independent field for the purposes of computing the quadratic action.  It is also simpler to replace the curvature-squared term  $K$ by an equivalent Lagrangian involving an auxiliary symmetric tensor field $f_{\mu\nu}$  \cite{Bergshoeff:2009hq}; the resulting action is
\bea\label{linstart}
I[g,f,S] &=&  \frac{1}{\kappa^2} \int d^3 x\, e\left[ \left(MS - 2\sigma S^2 + \frac{S^3}{\check\mu} - \frac{3S^4}{2(m^\prime)^2} \right) + \sigma R
+ \frac{RS}{2\check\mu} - \frac{RS^2}{2(m^\prime{}^\prime)^2} \right. \nonumber \\
&&\left.  \qquad \qquad  + \ f_{\mu\nu} G^{\mu\nu}-\frac{1}{2}m^2g^{\mu\nu}g^{\rho\sigma} f_{\mu[\rho}f_{\nu] \sigma}  \right] \, .
\eea
We wish to find the quadratic approximation to this action  in a supersymmetric  adS vacuum  with cosmological constant $\Lambda= -1/\ell^2$.


\subsection{Quadratic approximation}\label{quadratic}


We now set
\bea\label{fluctuations}
g_{\mu\nu} &=&  \bar{g}_{\mu\nu}+\kappa h_{\mu\nu}\, , \qquad S \ = \  \pm \ell^{-1}+\kappa s \, ,
\nonumber \\
f_{\mu\nu} &=& - \frac{1}{\ell^2m^2}\left[\bar{g}_{\mu\nu}+ \kappa
h_{\mu\nu} + \ell^2\kappa k_{\mu\nu}\right]\, ,
\eea
where $h_{\mu\nu}$, $k_{\mu\nu}$ and $s$ are independent fluctuation fields\footnote{The mass dimensions of these fluctuation fields are:  $[h] =\frac{1}{2}$, $[s]=\frac{3}{2}$ and
$[k]=\frac{5}{2}$.}, and  $\bar g_{\mu\nu}$ is the background adS  metric.
We shall use the notation $\bar D$ to indicate a covariant derivative with respect to the standard Levi-Civita connection for the background metric.  Expanding the full Ricci tensor about the adS background we find that
\be
R_{\mu\nu} \ = \  -2\ell^{-2} \bar g_{\mu\nu} +\kappa
R_{\mu\nu}^{(1)}+\kappa^2 R_{\mu\nu}^{(2)}+{\cal O}(\kappa^3)\;,
\ee
where
 \be\label{R1}
  R_{\mu\nu}^{(1)} \ = \ -\frac{1}{2}\left(\bar D^2
  h_{\mu\nu}- \bar D^{\rho}\bar D_{\mu}h_{\rho\nu}-\bar D^{\rho}\bar D_{\nu}h_{\rho\mu}
  + \bar D_{\mu}\bar D_{\nu}h\right)\;.
 \ee
We will need only the trace of the $\kappa^2$ term, which is
 \bea\label{Rsquare}
  \bar{g}^{\mu\nu}R_{\mu\nu}^{(2)} \ = \
  \frac{1}{2}h^{\mu\nu}\left(R_{\mu\nu}^{(1)}-\tfrac{1}{2}R^{(1)}\bar{g}_{\mu\nu}\right)+\text{total derivative}\, ,
 \eea
where $R^{(1)}$ is the trace of $R_{\mu\nu}^{(1)}$ in the background metric.

At this point it is useful to recall the gauge symmetries at the linearized level and what the gauge-invariant objects are. The metric fluctuation
transforms in the standard way under linearized diffeomorphisms,
 \bea\label{lingauge}
  \delta_{\xi}h_{\mu\nu} \ = \
  \bar D_{\mu}\xi_{\nu}+\bar D_{\nu}\xi_{\mu}\;,
 \eea
while $k_{\mu\nu}$ and $s$ have been defined such that they are gauge-invariant. The invariant curvature of $h_{\mu\nu}$ is
given by the linearized Einstein tensor modified by the cosmological constant,
 \bea\label{EinsteinDS}
 \begin{split}
  {\cal G}_{\mu\nu}(h) \ &\equiv \ G_{\mu\nu}^{(1)}(h)+\Lambda
  h_{\mu\nu} \\
  \ &= \ R_{\mu\nu}^{(1)}-\ft12 R^{(1)}\bar{g}_{\mu\nu}-2\Lambda
  h_{\mu\nu}+\Lambda h\bar{g}_{\mu\nu}\;,
 \end{split}
 \eea
which is the tensor that defines the linearized field equations of pure Einstein gravity with cosmological constant.

Expanding the action about the vacuum, we find that all terms linear in the fluctuations cancel provided
\bea\label{SUSYADS}
M = \pm \frac{4\sigma}{\ell}   \mp  \frac{2}{5 \check m^2 \ell^3}\, ,
\eea
which is the $S$ field equation in a supersymmetric vacuum with $\bar S =\pm\ell^{-1}$; this confirms the existence of these vacua. For the quadratic terms in the Lagrangian we find the  manifestly gauge-invariant expression
\bea\label{L2Total}
L^{(2)} &=& -\frac{1}{2}\hat\sigma
h^{\mu\nu}{\cal G}_{\mu\nu}(h) + \frac{a}{\ell m^2}  s\bar g^{\mu\nu}{\cal G}_{\mu\nu}(h)
-\frac{1}{ m^2}k^{\mu\nu}{\cal G}_{\mu\nu}(h)
\\  \nonumber
&&-\frac{1}{4m^2}\left(k^{\mu\nu}k_{\mu\nu}
-k^2\right) -\left(2\sigma  \mp \frac{3}{\ell \check \mu} + \frac{6}{\ell^2 m^2} - \frac{21}{5\ell^2 \check m^2}\right)s^2
\ ,
\eea
where $a$, the parameter defined in (\ref{adef}), is now given by
\be\label{coeffred} a =  -m^2 \left(\frac{\ell}{\check\mu}\mp
\frac{2}{(m^\prime{}^\prime)^2}\right)=
m^2\left(-\frac{\ell}{\check\mu} \pm\frac{2}{m^2} \mp
\frac{6}{5{\check m}^2}\right)\ . \ee In the present context, the
condition $a=0$ yields the quadratic approximation for  the
``generalized super-NMG'' case, and the two conditions $a=0$ and
$|\check\mu|=\infty$ yield the quadratic approximation to super-NMG.
As the analysis of propagating modes will depend crucially on
whether $a$ is zero or non-zero, and as the $a=0$ case is of more
relevance  to  ``massive gravity''. The parameter $\hat{\sigma}$
introduced in \eqref{sihat} will also play a significant role in
what follows; it is useful to note that this parameter may be
rewritten as
\be\label{altsighat}
\hat\sigma = \sigma + \frac{1}{2\ell^2 m^2}  \pm \frac{1}{4\ell\check\mu} \mp \frac{a}{4\ell^2 m^2} \;.
\ee

Our next goal  is  to analyze the modes propagated by  the  Lagrangian (\ref{L2Total}). After some field redefinitions, we will be able to do this by comparison with  Proca and Fierz-Pauli theory in anti de Sitter space. For the convenience of the reader,  we first review this topic; one of our aims will be to determine the bounds on the masses  of spin-$1$ and spin-$2$ particles in adS that are implied by the absence of tachyons.


\subsection{Review of Proca and Fierz-Pauli in adS}\label{FPreview}


For a vector field $A_{\mu}$ the massive Proca Lagrangian in an adS background is given by
 \bea
   {\cal L}_{\rm Proca} \ = \ -\frac{1}{4}F^{\mu\nu}F_{\mu\nu}-\frac{1}{2}{\cal M}^{2}A^{\mu}A_{\mu}\;.
 \eea
It propagates massive spin-1 modes;  in 3D this means that there are two modes, one of helicity $+1$ and one of helicity $-1$.
The existence of two modes can be seen by inspecting the field equations.
Variation with respect to $A_{\mu}$ yields
 \bea
   \bar{D}^{\mu}F_{\mu\nu}-{\cal M}^2 A_{\nu} \ = \ 0\qquad \Rightarrow \qquad \bar{D}^{\mu}A_{\mu} \ = \ 0\;,
 \eea
where the second equation (the subsidiary condition) follows by taking the divergence of the first equation.
The dynamical equation can then be written as
 \bea \label{dynproca}
  \left(\bar{D}^2-2\Lambda-{\cal M}^2\right)A_{\mu} \ = \ 0\;.
 \eea
The subsidiary condition yields one constraint, which implies that there are in total two propagating degrees of freedom.

For a symmetric tensor field $\varphi$ and mass parameter ${\cal
M}$, the FP Lagrangian in an adS background is \be L_{\rm FP}
(\varphi; {\cal M}^2) = -\frac12 \varphi^{\mu\nu}{\cal
G}_{\mu\nu}(\varphi) - \frac{1}{2}{\cal M}^2 \bar g^{\mu\nu}\bar
g^{\rho\sigma} \varphi_{\mu[\rho} \varphi_{\nu]\sigma}\, . \label {fpaction}\ee For
${\cal M}^2 \ne \Lambda$ this Lagrangian  propagates,
massive spin-2 modes;  in 3D this means that there are two modes,
one of helicity $+2$ and one of helicity $-2$. The presence of two
propagating degrees of freedom can be seen by inspecting the field
equation
\be\label{FPE}
{\cal G}_{\mu\nu}(\varphi)+\frac{1}{2}{\cal M}^2\left(\varphi_{\mu\nu}- \bar{g}_{\mu\nu} \bar\varphi\right) \ = \ 0\, , \qquad
\left(\bar\varphi \equiv \bar g^{\mu\nu} \varphi_{\mu\nu}\right).
\ee
Taking the divergence of this equation and using the Bianchi identity
$\bar{D}^{\mu}{\cal G}_{\mu\nu}=0$, we obtain
 \be\label{FPdiv}
  \bar{D}^{\mu}\varphi_{\mu\nu}-\bar{D}_{\nu}\bar\varphi  \ = \ 0 \qquad \Rightarrow \qquad
  \bar{D}^{\mu}\bar{D}^{\nu}\varphi_{\mu\nu}-\bar{D}^2\bar\varphi \ = \ 0\;.
 \ee
On the other hand, taking the trace of (\ref{FPE}) and using the explicit form of $R^{(1)}$ in (\ref{R1}), we get
 \be\label{hzero}
 \bar{D}^{\mu}\bar{D}^{\nu}\varphi_{\mu\nu}-\bar{D}^2\bar\varphi  =2\left(\Lambda-{\cal M}^2\right)\bar\varphi\, ,
 \ee
 where $\Lambda < 0$ is the cosmological constant.  Combining this with (\ref{FPdiv}) we conclude that $\bar\varphi=0$ provided that ${\cal M}^2 \ne \Lambda$ and hence that the symmetric tensor field $\varphi$ is subject to the subsidiary conditions
 \be\label{subFP}
 \bar{D}^{\mu}\varphi_{\mu\nu} \ = \ 0\;, \qquad \bar\varphi \ = \ 0\;.
 \ee
The remaining, dynamical, equation is
 \be\label{dynamical}
\left( \bar D^2 -2\Lambda -{\cal M}^2\right)\varphi_{\mu\nu} =0\, .
\ee
The subsidiary conditions impose $3+1$ constraints, so just  two degrees of freedom are propagated, and these can be shown to have helicities $\pm 2$. Observe that the specific Fierz-Pauli mass term is crucial to this result  because with a different relative coefficient it would not be possible to derive (\ref{FPdiv}), and the subsidiary condition  $\bar\varphi=0$, needed to eliminate scalar modes,  would not be a consequence of the field equations.

In the special case of ${\cal M}^2=\Lambda$, the FP field equation
does not imply that $\bar{\varphi} = 0$. In this case there is a
`hidden' gauge invariance,
\bea\label{hidden}
\delta_{\zeta}\bar{k}_{\mu\nu} \ = \ \bar{D}_{\mu}\bar{D}_{\nu}\zeta
+\Lambda\bar{g}_{\mu\nu}\zeta\;,
\eea
with scalar gauge parameter $\zeta$. This allows the trace $\bar\varphi$ to be set to zero by a gauge-fixing condition. Theories of this type are known as partially massless \cite{Deser:1983mm,Deser:2001us}, and in 3D they propagate a single mode without a well-defined helicity.

There is obviously a need for some lower bound on ${\cal M}^2$, in order
to avoid tachyons. Let us consider the generalization of
\eq{dynamical} to arbitrary integer spin $|s|$ \cite{Bianchi:2005ze}
\be\label{dprodimp}
  \left[\bar D^2 +
| s|\left(3-|s|\right)-\mathcal{M}^2\right]\varphi^{(s)}=0\ ,
\ee
where $\varphi^{(s)}$ denotes a traceless totally symmetric rank-$|s|$
tensor satisfying the `divergence-free' condition $\bar D^\mu
\varphi^{(s)}_{\mu\nu_1\dots \nu_{s-1}} =0$. For $|s|>0$, the action
from which this field equation is derived is gauge invariant when $\mathcal{M}^2=0$.
Expanding the field $\varphi^{(s)}$ in terms of the unitary
irreducible representations (UIRs) of the  adS$_3$  isometry group
$Sl(2;\bR)\times Sl(2; \bR)$, we find \cite{Deger:1998nm}
\be
{\bar D}^2 \varphi^{(s)}= E_0(E_0-2) -|s|\ ,
\label{eigenv}
\ee
where $(E_0,s)$ denotes the lowest weight UIR with lowest energy $E_0$
and helicity $s$. These UIRs are nonsingular at the origin and
normalizable with respect to the $SO(2,2)$ invariant measure
\cite{RLN,Breitenlohner:1982jf}. Using the above formula in
\eq{dprodimp}, we find
\be
\mathcal{M}^2 = \left(E_0 - |s| \right)(E_0 + |s| - 2)\ .
\ee
Now, it is well known that the unitarity of the representation with
lowest weight $(E_0,s)$ is given by \cite{Barut:1986dd}
\be
E_0 \ge |s| \ .
\ee
For $s=0$ we deduce that $\mathcal{M}^2\ge -1$, which is the 3D
version of the 4D Breitenlohner-Freedman bound
\cite{Breitenlohner:1982jf,Mezincescu:1984ev,HarunarRashid:1991bv}.
For $s\ge1$ we deduce that $\mathcal{M}^2\ge0$, as claimed for
$s=1,2$.


\subsection{Diagonalization}\label{diagonalization}


We are now ready to continue with our analysis of the quadratic Lagrangian (\ref{L2Total}). The results depend crucially on whether
the parameter $a$,  defined in \eq{coeffred},  is zero or non-zero, so we consider these cases separately.


\subsubsection{$a=0$}


When $a=0$ the field $s$ may be trivially eliminated and the
quadratic Lagrangian (\ref{L2Total}) reduces to \be\label{L2Total2}
L^{(2)} = -\frac{1}{2}\hat\sigma h^{\mu\nu}{\cal G}_{\mu\nu}(h)
-\frac{1}{ m^2}k^{\mu\nu}{\cal G}_{\mu\nu}(h)
-\frac{1}{4m^2}\left(k^{\mu\nu}k_{\mu\nu} -k^2\right)\,, \ee where
$\hat{\sigma}$ is the parameter of \eqref{altsighat}. This is
precisely eq.~(4.17) of \cite{Bergshoeff:2009aq} when
$|\check\mu|=\infty$, which corresponds to the super-NMG model; this
was to be expected because super-NMG has NMG as its bosonic
truncation.  The only difference between super-NMG and generalized
super-NMG in the context of a quadratic approximation is in the
definition of the parameter $\hat\sigma$. How we now proceed depends
on whether or not $\hat\sigma$ vanishes. We shall consider these two
subcases separately.

\begin{itemize}


\item  \underline{$\hat\sigma \ne 0:$}
%
%
In this case we define a new symmetric tensor fluctuation field
${\bar h}$ by
\be\label{fieldred}
h_{\mu\nu} \ = \ \bar{h}_{\mu\nu} - \frac{1}{m^2\hat\sigma}\, k_{\mu\nu}\ .
\ee
The quadratic Lagrangian then takes the diagonal form
\be\label{decolag} L^{(2)} =  -\frac{1}{2}\hat{\sigma}\,
\bar{h}^{\mu\nu}{\cal G}_{\mu\nu}(\bar{h}) - \frac{1}{m^4
\hat\sigma} \, L_{\rm FP}(k; -m^2\hat\sigma) \, ,\ee
where $L_{\mathrm{FP}}$ was defined in \eq{fpaction}. We see from
this result that $\hat\sigma$ has the interpretation as the effective EH coefficient in a non-Minkowski vacuum.
Because this term propagates no modes, we effectively have an FP Lagrangian with ${\cal M}^2= -m^2\hat\sigma$.
As we explained earlier, the absence of tachyons requires $\mathcal{M}^2\ge0$ (which is a stronger condition that
used in \cite{Bergshoeff:2009aq}) and hence $m^2\hat\sigma<0$. We also require $\hat\sigma<0$ for positive kinetic energy
(no ghosts) so we deduce that the combined conditions for no ghosts and no tachyons are
\be
m^2>0\, , \qquad \sigma + \frac{1}{2\ell^2 m^2}  \pm \frac{1}{4\ell\check\mu} <0\;.
\label{uc1}
\ee
Note that these conditions imply that $\sigma<0$ in the NMG limit $|\check\mu|\to\infty$, but
$\sigma>0$ is possible in the ``generalized'' case.

We should recall here that  the case  ${\cal M}^2=\Lambda$ is special because it corresponds
to a partially massless mode \cite{Deser:2001us}. It is not clear to us whether our  earlier conclusion that
$\mathcal{M}^2\ge0$ is required for the absence of tachyons also  applies  in this special case.


\item \underline{$\hat\sigma = 0:$}
%
%

In this special case, we see from (\ref{L2Total2}) that  the fluctuation field $h_{\mu\nu}$ is a Lagrange multiplier; the  constraint it imposes has
the general solution
\be\label{Vector}
k_{\mu\nu} =  2 \bar D_{(\mu} A_{\nu)}\, ,
\ee
for arbitrary vector field $A_\mu$. Using this solution we arrive at the equivalent Lagrangian
\be \label{ProcaLagrangean}
L^{(2)}  = - \frac{1}{4m^2} F^{\mu\nu}F_{\mu\nu} - \frac{2}{\ell^2 m^2} A^\mu A_\mu \, ,
\ee
where we have discarded a total derivative. This is a Proca Lagrangian for $A_\mu$, with positive kinetic energy provided
$m^2>0$ and  a specific value for  the mass.

Alternatively, the Proca equations may be deduced from the equations of motion of
(\ref{L2Total2}). The $k$ field equation is
\be\label{keq}
\mathcal{G}_{\mu \nu}(h) + \frac{1}{2} \left( k_{\mu \nu} - k \bar
g_{\mu \nu}\right)=0 \,.
\ee
When combined with the  Bianchi identity $\bar{D}^{\mu}{\cal G}_{\mu\nu}=0$ and the $h$ field equation (\ref{Vector}), this implies the
Proca equations that follow from (\ref{ProcaLagrangean}).  Provided $m^2>0$ these equations propagate  non-tachyonic modes of helicity $\pm1$.
This is consistent with the corresponding result for NMG \cite{Bergshoeff:2009aq}; however, whereas $\hat\sigma=0$ was there found to imply
$\sigma<0$, this is not true in the ``generalized'' case since  it follows from (\ref{altsighat})  that $\hat\sigma=0$ and  $a=0$ imply
\be
\sigma= - \frac{1}{2\ell^2 m^2} \mp \frac{1}{4\ell\check\mu}\ ,
\label{solvesig}
\ee
and this allows $\sigma<0$ when $\check\mu$ is finite.

Finally, we remark that  the equation (\ref{keq}) does not propagate any modes if one adopts the standard
Brown-Henneaux boundary conditions for the metric \cite{Brown:1986nw}  but weaker boundary conditions
allow well known logarithmic bulk modes \cite{Grumiller:2008qz}.  It may be verified that the Proca modes mentioned above are mapped by
\eq{Vector} into the first descendants of the logarithmic modes;  see \cite{Giribet:2008bw} for a detailed description of
precisely such a descendant mode.

The occurrence of different formulations at the linearized level is
similar  to what happens in TMG, in which case there exists a map of
the linearized field equation at the chiral point to that of a
topologically massive photon \cite{Carlip:2008jk}. Alternatively,
the linearized theory can be mapped, non-covariantly, to a scalar
field satisfying the Breitenlohner-Freedman bound
\cite{Carlip:2009ey}. The linearized solution of these equations are
related to the logarithmic solutions of the metric formulation, as
has been shown in \cite{Carlip:2009ey} for the scalar
parametrization.

\end{itemize}


\subsubsection{$a \ne 0$}


When $a\ne0$ we must return to the quadratic Lagrangian (\ref{L2Total}).  Again we must distinguish between the $\hat\sigma \ne0$ and the $\hat\sigma=0$ cases, so we consider them in turn.

\begin{itemize}


\item  \underline{$\hat\sigma \ne 0:$}
%
%
When $\hat\sigma\ne0$  the Lagrangian  becomes diagonal in terms of the  new symmetric tensor fluctuation fields $(\bar h, \bar k)$, defined by
\be\label{fieldred}
h_{\mu\nu} \ = \ \bar{h}_{\mu\nu} - \frac{1}{m^2\hat\sigma}\,  \bar{k}_{\mu\nu}\;, \qquad
k_{\mu\nu} \ = \ \bar{k}_{\mu\nu}+ a\ell^{-1}  s\, \bar{g}_{\mu\nu}\;,
\ee
where $a$ is the mass parameter defined  in (\ref{coeffred}). The quadratic Lagrangian then takes the form
\be\label{decolag}
L^{(2)} =  -\frac{1}{2}\hat{\sigma}\, \bar{h}^{\mu\nu}{\cal G}_{\mu\nu}(\bar{h})
- \frac{1}{m^4 \hat\sigma} \, L_{\rm FP}(\bar{k}; -m^2\hat\sigma) +  \frac{a}{\ell m^2} \bar k s - b s^2\;,
\ee
where
\be\label{bb}
b =  \frac{1}{2}\hat\sigma - \frac{\left(3a\mp2\right)\left(a\mp2\right)}{2\ell^2 m^2}\, .
\ee
If $b\ne 0$ then $s$ may be trivially eliminated; this will
give rise to an additional $\bar k^2$ mass term which will lead to a
non-unitary theory (since the specific FP mass term is crucial for
unitarity).

If $b=0$ then the field  $s$ becomes a Lagrange multiplier for the constraint  $\bar{k}=0$, which is one of the subsidiary conditions of the FP equations. However, the $\bar{k}_{\mu\nu}$ field equation now reads
\be\label{modFP}
{\cal G}_{\mu\nu}(\bar{k}) = m^2\hat\sigma \left(\ft12\bar k_{\mu\nu}
-  a \ell^{-1}s\, \bar g_{\mu\nu} \right)\;.
\ee
Taking the divergence we deduce that
\be\label{divcond}
\bar D^\mu \bar k_{\mu\nu} = 2 a \ell^{-1} \partial_\nu s \quad \Rightarrow \quad \bar D^\mu\bar D^\nu \bar k_{\mu\nu} = 2 a\ell^{-1} \bar D^2 s\, .
\ee
Taking the trace, one finds
\be
 \bar{D}^\mu\bar D^\nu \bar k_{\mu\nu} = 6 \ell^{-1} m^2\hat\sigma a s\;,
 \ee
 and in combination with (\ref{divcond}) this gives
 \be
 \left(\bar D^2 - 3m^2\hat \sigma\right) s =0\, .
 \ee
 In other words, the fluctuation $s$ about the vacuum value of $S$ is now a propagating mode! Whether the theory is ghost-free in presence of this mode, however, remains to be investigated.

 The fact that the `auxiliary' field propagates is surprising in view of the fact that the field equation for $S$ is algebraic, a cubic equation in fact, but the coefficients of this cubic equation are not constants when $a\ne0$. We earlier argued that one may solve for $S$ as a power series in $R$ in this case, but all orders of this series are relevant to an expansion about adS, so it is not guaranteed that the solutions for fluctuations of $S$ will be local functions  of the coefficients.
We now see that  an `auxiliary $S$', in the generalized sense that
we have permitted in this section, is not equivalent to
`non-propagating $S$'.


\item  \underline{$\hat\sigma = 0:$}
%
%
Setting  $\hat\sigma=0$ in the Lagrangian \eq{L2Total}, but now allowing for $a\ne0$, we find that the fluctuation field $h_{\mu\nu}$ becomes a Lagrange multiplier, as before. The  constraint it imposes has the general solution
\be
k_{\mu\nu} =  a\ell^{-1} s {\bar g}_{\mu\nu} + 2 \bar D_{(\mu} A_{\nu)}\, ,
\ee
for arbitrary vector field $A_\mu$. Using this solution we arrive at the equivalent Lagrangian
\be
L^{(2)}  = - \frac{1}{4m^2} F^{\mu\nu}F_{\mu\nu} - \frac{2}{\ell^2 m^2} A^\mu A_\mu +\frac{2a}{\ell m^2} s {\bar D}^\mu A_\mu - b  s^2\ ,
\ee
where we now have
\be
b = - \frac{\left(3a\mp2\right)\left(a\mp2\right)}{2\ell^2 m^2}\
\ee
If $b \ne 0$, then the field $s$ can be trivially eliminated, as before,  and this will give rise to additional $({\bar D}\cdot A)^2$ terms which will lead to a non-unitary theory (since the standard Proca form of the action is needed for unitarity).

If $b=0$, then the field $s$ becomes a Lagrange multiplier for the constraint $\bar D\cdot A=0$, which is the Proca subsidiary condition. Furthermore, the Proca equation is now modified to
\be
\bar D^\mu F_{\mu\nu} - 4\ell^{-2}A_\nu = 2 a\ell^{-1} \partial_\nu s\;.
\ee
Taking the divergence of this equation we deduce that
\be
\bar D^2 s=0\ .
\ee
So the fluctuation field $s$ propagates a scalar mode. The unitarity of the model in presence of this mode remains to be investigated.

\end{itemize}

\subsection{Summary}

A curiosity that our analysis has uncovered is that  a scalar field may be  ``auxiliary'' in the sense of having no kinetic term but still propagate modes in a non-Minkowski vacuum if it is coupled to scalar products of propagating fields.  The distinction between ``auxiliary'' and ``non-propagating'' boils down, in the cases analysed, to whether a dimensionless parameter $a$ is non-zero (the generic case) or zero (the ``non-propagating'' case). The latter option yields the cases that we have referred to as those of ``generalized super-NMG''.  The more general ``auxiliary S'' models, with $a\ne0$, propagate scalar modes and are generically non-unitary although there may be special subcases that are perturbatively unitary.

Within ``generalized super-GMG'' we find the ``super-NMG'' models. Since these have NMG as a bosonic truncation (albeit with a restricted range of the NMG parameters) we should expect agreement with the results found for NMG in  \cite{Bergshoeff:2009aq} . We do, except  for the stricter condition on perturbative unitarity that follows from  the stronger bound on the spin-$2$ Fierz-Pauli  mass in adS vacua that we have justified here.

We have also shown that  the super-NMG results extend to  ``generalized super-NMG'', the only difference being that the ``effective'' EH coefficient $\hat\sigma$ now depends on an additional parameter. This allows perturbative unitarity to be made consistent with $\sigma>0$, i.e. with  ``right-sign'' EH term in the action. However, it should be recalled that ``generalized super-NMG'' is not actually a class of ``models'' because its definition depends on a choice of adS vacuum; in particular, the conclusion that $\sigma<0$ is needed for perturbative unitarity in Minkowski vacua is unchanged.


\section{Discussion}\label{sec:conclusions}
\setcounter{equation}{0}


In this paper we have completed a study of three-dimensional (3D) ${\cal N}=1$ supergravity theories with generic curvature-squared terms that was
begun in  \cite{Andringa:2009yc}. That paper was titled ``Massive 3D supergravity'' but contact was made with the massive gravity models introduced in
\cite{Bergshoeff:2009hq} only in the context of an  expansion about Minkowski spacetime, where non-linear features are not crucial.  The space of
non-Minkowski vacua found in  \cite{Andringa:2009yc} had no obvious relation to the space of non-Minkowski vacua found in \cite{Bergshoeff:2009hq},
and neither did there appear to be any supergravity model with a bosonic truncation that could be identified with a massive gravity model. As we said
in the introduction, these unsatisfactory features suggest that  there is some ingredient missing from  the analysis of \cite{Andringa:2009yc}, and we
have shown here that this is indeed the case. The supergravity results of  \cite{Andringa:2009yc} are correct but incomplete because there is an
additional super-invariant  involving the auxiliary scalar field $S$ of  ${\cal N}=1$ supergravity that contributes to the terms with the {\it
dimension} of curvature-squared terms but not to the curvature-squared terms themselves. Incorporating this invariant into a more general action
allows the choice of a special case in which $S$  can be eliminated, at least classically, to yield a model that is identical to the `cosmological'
extension of  the ``general massive gravity'' (GMG) model introduced in  \cite{Bergshoeff:2009hq}, and this includes as a special case the
`cosmological' extension of the parity-preserving ``new massive gravity'' (NMG) model  studied in detail in \cite{Bergshoeff:2009aq}.

Actually, it is overstating the case to say that the new results of this paper  are suggested by complications for non-Minkowski found in
\cite{Andringa:2009yc}  because it is far from obvious, a priori, that a higher-derivative gravity model {\it should} arise as the truncation of a
supergravity model. In fact,  the results of this paper confirm the contrary conclusion for generic curvature-squared models, since the supergravity
extension of the generic model necessarily involves a kinetic term for the `auxiliary' scalar, thus propagating a field that was not present
initially. The special feature of the NMG and GMG models, already noted in \cite{Bergshoeff:2009hq}, is that this term is absent, so that the
`auxiliary' field $S$ really does remain auxiliary in the sense that its field equation is algebraic, in fact cubic. However, the incomplete results
of  \cite{Andringa:2009yc} led to the conclusion that  this cubic equation necessarily has coefficients that are not all constant but depend upon the
scalar curvature $R$. Elimination of $S$ then leads to an additional power series in $R$ contribution to the action; in particular, it leads to an
additional unwanted $R^2$ term. Had this been  the last word on the matter, it would have encouraged the view that that massive 3D gravities are mere
curiosities. Conversely, the fact that  one can recover NMG or GMG as bosonic truncations of a 3D supergravity model, as shown in this paper, paves
the way to a further study of extended super-GMG models and encourages the belief that these models should have a role to play in some `bigger
picture'.

The main point of interest in the new massive gravity models such as NMG and GMG is the
fact that the higher-derivative terms are consistent with unitarity, at least in the
Minkowski vacuum. This result was shown in \cite{Andringa:2009yc} to extend to the spin-$\tfrac{3}{2}$ sector of the
supergravity models, as is of course implied by supersymmetry. The issue of unitarity in
adS vacua was studied in detail in \cite{Bergshoeff:2009aq} for NMG and we have here extended this analysis
to super-NMG and some variants of it that also preserve parity. As the bosonic truncation
of super-NMG is equivalent to NMG after elimination of the supergravity auxiliary field
$S$, and as all adS vacua of NMG correspond to a supersymmetric adS vacuum of super-NMG,
the results of \cite{Bergshoeff:2009aq} for linearization about an adS vacuum extend immediately to
linearization of super-NMG about a supersymmetric adS vacuum; in particular, there is no
need to consider the spin-$\tfrac{3}{2}$ sector because this is determined by supersymmetry in a
supersymmetric vacuum.

There is one caveat: we have shown here that the Fierz-Pauli mass $\mathcal{M}$ for a spin-$2$ field in adS$_3$
must satisfy $\mathcal{M}^2\ge0$ in order that the associated spin-$2$ particle not be a tachyon\footnote{We presume that this result is known but there are suggestions in the literature of a ``Breitenlohner-Freedman bound for spin $2$'' that allows $\mathcal{M}^2\ge-1$, as for spin zero.} whereas we
allowed (provisionally) for a weaker bound  in \cite{Bergshoeff:2009aq}. This means that the  range of parameters for which the linearized theory is perturbatively unitary  is more restricted than stated in \cite{Bergshoeff:2009aq}.
Another subtlety is that although  super-NMG has been defined as the model as for which
the bosonic truncation yields NMG after elimination of the auxiliary field $S$, there is a larger class of models
for which the field linearized equations coincide with those of NMG  if the parameters are tuned to the choice of
vacuum; specifically, we can tune the parameters so that the field equation for the
fluctuation of $S$ is algebraic. In this way, we slightly enlarge the class of models
that are perturbatively  unitary in an adS vacuum. Within this larger class perturbative unitarity is consistent with
either sign of the Einstein-Hilbert term provided the new parameter $\check{\mu}$ introduced in (\ref{genI}) is chosen appropriately.

Finally, we briefly consider the two-dimensional CFTs that might be
holographically related to the massive gravity models above when expanded about a
supersymmetric  adS vacuum.  Actually, we should expect a holographically dual
{\it superconformal} field theory, i.e. an SCFT, but it is unclear to us how the fermions
may be taken into account in a semi-classical approximation to the bulk supergravity theory,
so we instead consider only the bosonic truncations.
According to the Brown-Henneaux analysis, for generic adS$_3$
gravity theories the asymptotic symmetry group consists of two
copies of the Virasoro algebra corresponding to the two-dimensional
conformal symmetry \cite{Brown:1986nw}. Their central charges encode
important information about unitarity and the entropy of BTZ black
holes. In the case of parity-preserving gravity theories that
contain higher powers of the curvature tensor the (left and right) central charges
are given by  \cite{Saida:1999ec,Kraus:2005vz}
\bea c_{L}=c_{R}=\frac{\ell}{2G_3}g_{\mu\nu}\frac{\partial{\cal
L}_{3}}{\partial R_{\mu\nu}}\;. \eea
In the presence of the Lorentz Chern-Simons term, we need to add the
contributions $\pm3/(2G_3\mu)$ to $c_{L,R}$ \cite{Kraus:2005zm}. Taking all these
considerations into account, and starting from the general model
(\ref{newlagr}), we  obtain the following values for the central charges
\bea\label{central} c_{L,R}  \ =  \
\frac{3\ell}{2G_3}\left(\sigma+\frac{3}{10(\check{m})^2\ell^2}
\pm\frac{1}{2\check{\mu}\ell}\pm\frac{1}{\mu\ell}\right) \ \equiv \
\frac{3\ell}{2G_3}\left({\hat\sigma}\pm \frac{1}{\mu\ell}\right)
\;, \eea
where we used $\Lambda=-1/\ell^2$.  In the special case of super-GMG we have
\be
\hat\sigma= \sigma+ \frac{1}{2\ell^2 m^2}\, ,
\ee
and hence agreement with the results of \cite{Andringa:2009yc}. Perhaps the most significant
feature of the formula (\ref{central}) is that $\hat\sigma$ is the parameter determining the sign of the {\it effective}
linearized EH term in the chosen adS background, which  must be negative for perturbative unitarity. This means
that the difficulty encountered in all previous massive 3D gravity models, that one must choose between
non-unitary gravitons or  negative mass BTZ black holes,  is a rather general one that is not resolved in supergravity, no matter
how one adjusts the parameters.

\subsection*{Acknowledgments}

We acknowledge helpful discussions and correspondence with Steve
Carlip, Geoffrey Comp\`ere, Daniel Grumiller and  Paul Howe. We also
thank each others home institutions for the hospitality extended
during visits. PKT is supported by an EPSRC Senior Fellowship. The
research of ES is supported in part by NSF grants PHY-0555575 and
PHY-0906222.  The work of O.H. is supported by the DFG--The German
Science Foundation and in part by funds provided by the U.S.
Department of Energy (DOE) under the cooperative research agreement
DE-FG02-05ER41360.

\newpage






\end{document}